%% file: DST_habib_v3_edited_letter_form.tex
\documentclass[aps,pre,twocolumn,twoside,preprintnumbers,amsmath,amssymb,longbibliography]{revtex4-1}

\usepackage{graphicx}
\usepackage{dcolumn}
\usepackage{bm}
\usepackage{placeins}
\usepackage{amsmath,amsfonts,amssymb}
\usepackage{amssymb}
\usepackage{bbm}
\usepackage{mathrsfs}
\usepackage[sans]{dsfont}
\usepackage{pifont}
\usepackage{tabularx}

\usepackage{amsmath,amsfonts,amssymb}
\usepackage[usenames]{color}

\usepackage{color}
\usepackage[dvipsnames]{xcolor}
\usepackage{float}
\usepackage{enumerate}

\newcommand{\gd}{\ensuremath{\dot\gamma}}
\newcommand{\ie}{{ \it i.e.,~}}

\newcommand{\etal}{{ \it et al.~}}

\newcommand{\sxy}{\ensuremath{\sigma_{xy}}}
\newcommand{\sxyn}{\ensuremath{\sigma_{xy,n}}}
\newcommand{\sxyt}{\ensuremath{\sigma_{xy,t}}}
\newcommand{\syxt}{\ensuremath{\sigma_{yx,t}}}

\newcommand{\pst}{\ensuremath{P(\delta\sigma_{xy}^-,\gd^+)}}
\newcommand{\pvt}{\ensuremath{P(\gd^-, \delta\sigma_{xy}^+)}}

\newcommand{\beq}{\begin{equation}}
\newcommand{\eeq}{\end{equation}}

\begin{document}


\title{Origin of Spatiotemporal Fluctuations in Discontinuous Shear Thickening}

\author{S. H. E. Rahbari$^1$} \email{habib.rahbari@gmail.com}


\affiliation{$^1$ School of Physics, Korea Institute for Advanced
  Study, Seoul 02455, Korea}

\begin{abstract}
  
Rheological phase transitions open the door to the less explored realm of non-equilibrium phase transitions. The main mechanism driving  these transitions is usually mechanical perturbation by shear--- an unjamming mechanism. Investigating discontinuous shear thickening (DST) is  challenging because the shear counterintuitively  acts as a jamming mechanism. Moreover, at the brink of this transition, a thickening material exhibits fluctuations that extend both spatially and temporally. Despite recent extensive research, the origins of such spatiotemporal fluctuations remain unidentified. Here, we investigate large fluctuations in DST by using versatile tools of stochastic thermodynamics. We discover a non-equilibrium dichotomy in the underlying mechanisms that give rise to large fluctuations and demonstrate that this dichotomy is a manifestation of novel collective behaviors across the transition. We then reveal the origin of spatiotemporal fluctuations in the shear thickening transition. Our study emphasizes the roles of stochastic thermodynamics tools in investigating non-equilibrium phase transitions, and demonstrates that these transitions are accompanied by simple dichotomies. We expect that our general approach  will pave the way to unmasking the nature of non-equilibrium phase transitions. 
  
\end{abstract}




\maketitle

{\bf Introduction.} Discontinuous shear thickening (DST),  an abrupt shear-driven fluid-solid transition, occurs in a wide range of soft material systems such as Brownian and non-Brownian suspensions as well as granular materials~\cite{brown_2014}. In suspensions, the transition is the result of an interplay between a stabilizing mechanism and frictional contacts due to the roughness of particles. At low flow speeds, the stabilizing  mechanism keeps particles apart. This mechanism  is usually a repulsive interaction  such as steric forces and/or electrostatic repulsion of double layers that introduces a threshold stress. At high flow speeds, when the threshold stress is overcome by the shear forces, a proliferation of frictional contacts results in an abrupt increase of more than an order of magnitude of viscosity--- the hallmark of a DST~\cite{morris_2009}. The stabilizing mechanism is system dependent; however, the frictional contacts are  essential for DST~\cite{seto_2013}. This behavior is well captured by the Wyart-Cates model, which describes DST as a transition from frictionless  to frictional rheologies, due to the proliferation of frictional contacts~\cite{wyart_2014}. In granular materials, DST is believed to be a result of frustration of the tendency of granular systems to dilate under shear~\cite{otsuki_2011, otsuki_2018}. We stress that the canonical mechanism responsible for DST in all soft material systems undergoing DST is inter-particle friction.  Notably, the friction is a (non-Brownian) granular mechanism~\cite{kawasaki_2018}. \\


Despite different mechanisms underlying DST in all the aforementioned soft material systems, DST has one consistent aspect: near thickening transition a system becomes unstable with spatiotemporal fluctuations.
In suspensions, temporal fluctuations appear as oscillations and chaotic time series akin to turbulence~\cite{saint_2018}, an effect dubbed rheochaos~\cite{cates_2002}. Moreover, spatial fluctuations  result in intermittent stress heterogeneities. These stress anomalies propagate along the vorticity direction, and they are referred to as vorticity bands~\cite{olmsted_2008, chacko_2018}. The strength of these fluctuations is enhanced with increasing stress, as confirmed by recent experiments using advanced techniques for the measurement of local rheology~\cite{rathee_2017, saint_2018}. Similar spatiotemporal fluctuations  have also been reported in extensive detail for frictional granular materials undergoing DST, in a series of publications by Heussinger, Zippelius and co-workers~\cite{grob_2014, grob_2016, saw_2019}.\\

These spatiotemporal fluctuations are believed to be the precursors of the shear thickening transition, yet the origin of these fluctuations remains a mystery. One reason for this lack of understanding is that, similarly to the glass transition~\cite{holmes_2005}, usual measures, such as pair correlation functions, show no signature of any dramatic change in micro-physics across DST~\cite{thomas_2018}. Here, we reveal the origin of the spatiotemporal fluctuations by using a stochastic thermodynamic approach~\cite{seifert_2012, sekimoto_2010} that has been specifically adopted for rheological phase transitions~\cite{rahbari_2017}. Stochastic thermodynamics is a powerful tool enabling thermodynamic quantities such as work, heat and power to be defined at the mesoscopic scale.\\

{\bf Results.} Here, we perform two dimensional molecular dynamics simulations of bidisperse frictional disks in a simple shear flow. We neglect thermal and hydrodynamic forces to focus on the central role of frictional forces in DST. However, a generalization of our approach to include hydrodynamic and thermal effects would be straightforward. Details of the simulations are given in the Methods of Supplementary Information (SI). A typical flow curve of the system is given
in Supplementary Fig.~1-a, where the stress abruptly changes
by more than one order of magnitude at a critical shear rate $\gd_c \simeq 10^{-5}$. The critical shear rate separates two
distinct fluid and solid states. We demonstrate that for $\gd<\gd_c$, the normal component of the shear stress is dominant; however, above $\gd_c$, the tangential component predominates in the momentum transfer. This result once again emphasizes the fundamental role played by friction in DST.  Most of physical measures such as pressure (Supplementary Fig. 2), coordination number (Supplementary Fig. 3) faithfully present the inherent state of the system consistent with the rheology. However, the kinetic temperature of the system 
(Supplementary  Fig.~1-b), known as the granular temperature $T_G$, has misleading behavior for $\gd> \gd_c$. Therefore, the kinetic temperature cannot be used as a real measure of fluctuations in this system, and an alternative measure is required. To resolve this
discrepancy, an effective temperature that complies with the rheology is required. To this end, we investigate large fluctuations in injected power $p$ by using the tools of stochastic thermodynamics to derive the effective temperature. \\

\begin{figure*}[ht]
	\centering
        \hfill\includegraphics[width=0.54\textwidth]{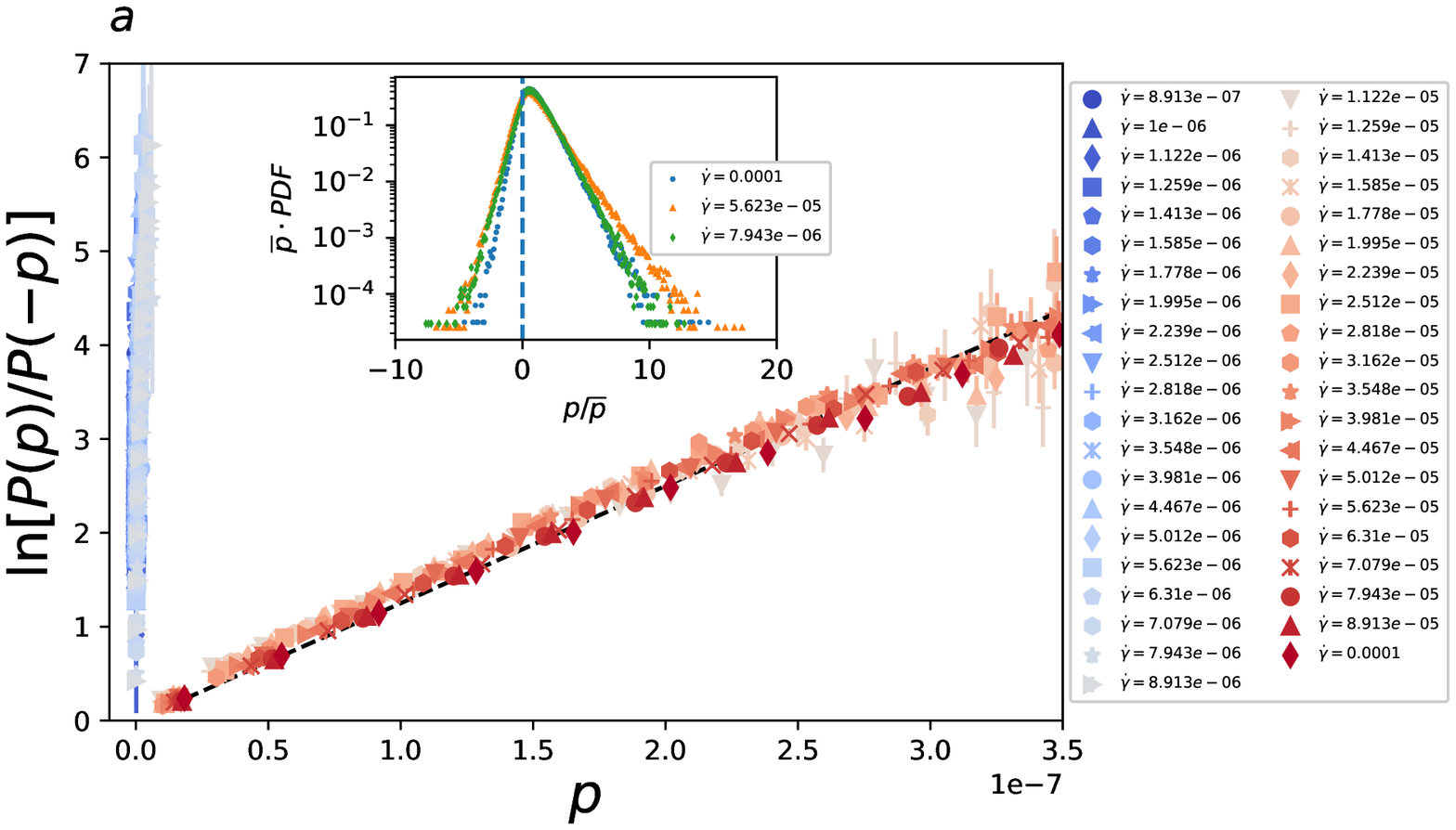}
        \hfill\includegraphics[width=0.44\textwidth]{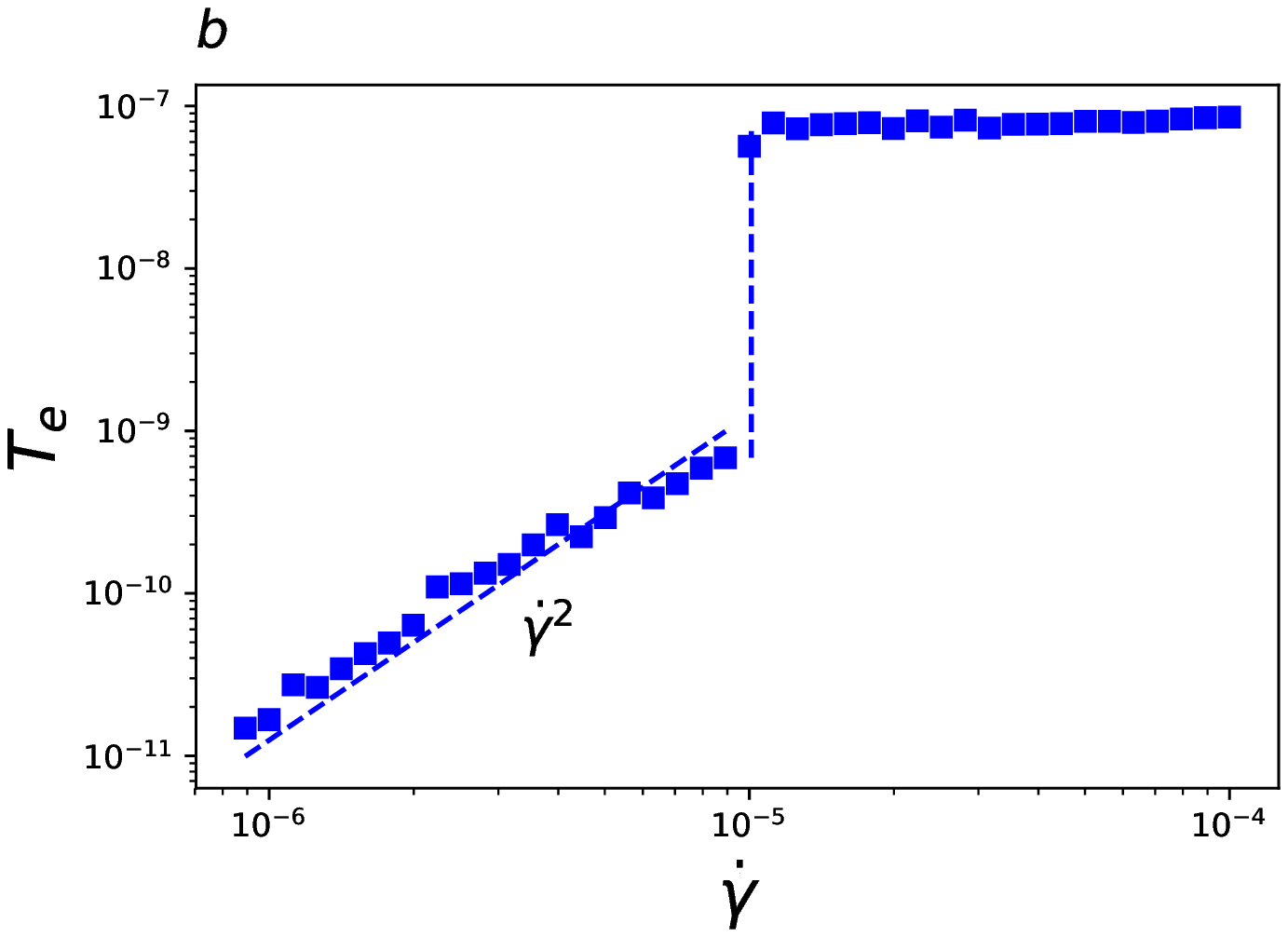}
	\hfill\\
	\caption[]{{\bf Fluctuation relation of power.} (a) The inset shows the PDF of power for various shear rates. It displays a universal double-exponential Boltzmann-type distribution. In the main panel, we examine the fluctuation relation proposed in Eq.\ref{eq:ft}, which is successfully verified by our data. (b) The effective temperature $T_e$ as a function of the shear rate $\gd$ is plotted. Bagnold behavior is found in the fluid-branch, which crosses over to a rate-independent behavior in the solid-branch, in accordance with the rheology. The effective temperature increases almost two orders of magnitude after thickening. The packing fraction is $\phi = 0.81$, and the number of particles is $N = 16384$. 
		\label{fig:tot_power_FT}}
\end{figure*}

The injected power due to simple shear is consumed at both translational and rotational degrees of freedom. Thus, we define the local injected power as:
\begin{equation}
p = \sxyn \gd + \delta\sxy \gd,
\label{eq:power}
\end{equation}
where $\gd$ is the local shear rate, $\sxyn$ is the normal component of the local shear stress, and $\delta \sxy =\sxyt -\syxt $ is the couple-stress, which is equal to difference in the off-diagonal components of the tangential part of the shear stress (the normal component cancels out, because $\sxyn = \sigma_{yx, n}$). The power is computed locally in the rectangle  of the length of the system size and width of $w = 2R$, with $R = 0.7$ radius of larger particles, along the shearing direction. We cross-checked our results for a wider bin of width $w = 4R$ and obtained similar results. A typical probability distribution function (PDF) of power is shown in the inset in Fig.\ref{fig:tot_power_FT}-a. The distribution is a double-exponential  Boltzmann-type distribution. Notably, such a Boltzmann-type PDF of stochastic thermodynamics quantities has been reported for various far-from equilibrium systems, for example, the PDF of work done on particles in a frictional granular system under pure shear~\cite{zheng_2018_2}, injected power in turbulence~\cite{bos_2019}, power in frictionless disks under shear~\cite{rahbari_2017}, and some cases reported by Gerloff and Klapp on the confined colloidal suspensions in shear flow~\cite{gerloff_2018}. Thus, the PDF of stochastic thermodynamic quantities has some common features across various non-equilibrium systems that have been overlooked to date. A power equal to $p$ given by Eq.~\ref{eq:power} is dissipated in a sub-system. This is akin to entropy production in a thermodynamic system. However, owing to a large fluctuation, the sub-system can give up the power, akin to entropy consumption~\cite{seifert_2012}. As shown in the inset, there exists a substantial part for large fluctuations in power given by $p<0$. We examine an instantaneous detailed fluctuation relation comparing the ratio of the PDF of the entropy production rate $\mathcal{P}(p)$ to that of the entropy consumption rate $\mathcal{P}(-p)$ via:
\begin{equation}
\ln \frac{\mathcal{P}(p)}{\mathcal{P}(-p)} = \beta p,
\label{eq:ft}
\end{equation}
in which $1/\beta =  T_e / \tau_e$ is the ratio of  an effective temperature to a time-scale. In the main panel of Fig.\ref{fig:tot_power_FT}-a, we plot this ratio for various shear rates. A linear dependence is clearly recovered, and thus the fluctuation relation is verified by our data. For $\gd< \gd_c$ (the blue data), we observe a large slope reflecting a very small effective temperature in the fluid phase. Although not visible in this regime, the slope slightly changes with shear rate. For $\gd>\gd_c$, the data show a substantially smaller slope, thus implying a considerably larger $T_e$. Moreover, in the solid-branch, data of various shear rates superimpose, and  the slope becomes independent of the shear rate, in agreement with the rheology. To provide a clear demonstration, in Fig.\ref{fig:tot_power_FT}-b we display the effective temperature $T_e$ versus the shear rate. The time scale $\tau_e$ is related to repulsive forces, because dissipative forces are negligible at very small shear rates. The effective temperature is computed via a direct linear regression by using Eq.~\ref{eq:ft}. Whereas a Bagnold dependence is obtained in the fluid-branch, $T_e$ resembles the behavior of the shear stress in the solid branch. Therefore, our proposed fluctuation relation gives rise to an effective temperature that behaves consistently with the rheology. Whereas the kinetic temperature shows an order of magnitude increase in fluctuations after thickening, the effective temperature increases more than two orders of magnitude. An effective temperature is a thermodynamic tool that enables mapping of a non-equilibrium state to an equilibrium thermodynamic state~\cite{makse_2002}. Therefore, our measured effective temperature $T_e$ may be interpreted as follows: DST may be mapped to an equilibrium transition from a finite temperature to an infinite temperature with large fluctuations. We will show later that this case is true for DST. Because the contribution of frictional forces predominates the rheology, we now focus on large fluctuations due to  frictional forces. \\

\begin{figure}[b]
	\begin{center}
	\hfill\includegraphics[width=.5\textwidth]{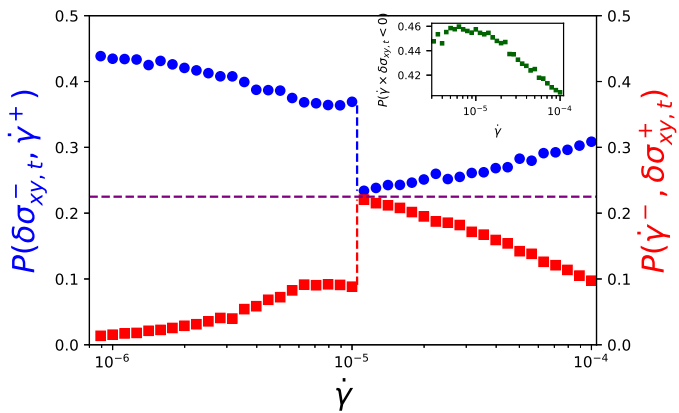}
	\hfill\\
	\caption[]{{\bf Decomposition of probabilities.} The inset shows $P(\delta\sxy  \gd < 0)$  as a function of the shear rate. In the solid phase, $P(\delta\sxy  \gd < 0)$ decreases monotonically with increasing shear rate. No discontinuity is observed at the critical shear rate. Main figure: joint probabilities $\pst$ and  $\pvt$ are displayed by the blue circles and red squares, respectively. Notably, a decomposition of joint probabilities into a mirror-image dichotomy has been discovered. Whereas $ \pst$ is reduced almost twice at $\gd_c$ and then increases as a function of the shear rate,  $\pvt$ is enhanced by the same rate, and it decreases with the shear rate in the fluid phase. Interestingly at $\gd = \gd_c$, $\pst= \pvt$. The packing fraction is $\phi = 0.81$, and the number of particles is $N = 16384$.
		\label{fig:neg_power_t}}
	\end{center}
\end{figure}

{\bf Large fluctuations due to frictional forces.} A large fluctuation caused by frictional forces occurs when the second term of the right-hand-side (rhs) of Eq.~\ref{eq:power} becomes negative, \ie $p_t = \delta\sxy \times \gd <0 $. In the Fig.~\ref{fig:neg_power_t}-a inset, the probability for such a large fluctuation  $P(\delta\sxy  \gd < 0)$ is plotted versus the shear rate. $P(\delta\sxy  \gd < 0)$ is almost constant (within errorbars) in the fluid branch, and it decreases monotonically in the solid branch. Because $p_t $ is a product of two terms, the couple-stress $\delta\sxy$ and the shear rate $\gd$, a negative $p_t$ can be  due to  either $\delta\sxy < 0$ with $\gd>0$  or 
$\gd<0$ with $\delta\sxy > 0$. This can be mathematically expressed as a decomposition relation
\begin{equation}
P(\delta\sxy  \gd< 0) = \pst + \pvt,
\label{eq:joint_p_t}
\end{equation}
in which $P( , )$  is a joint probability. In the main panel of Fig.~\ref{fig:neg_power_t}, we display $P(\delta\sxy ^-, \gd^+)$ (left Y-axis) and $P(\gd^-, \delta\sxy^+)$ (right Y-axis) with blue circles and red squares, respectively. A dazzling pattern emerges, indicating a decomposition of $P(\delta\sxy  \gd < 0)$ into two distinct branches for $\pst$ and $\pvt$. Moreover, this decomposition has two unique features. First, $\pst$ and $\pvt$ are approximately mirror images of one another, and second, in both solid and fluid states, their dependence on $\gd$ is dichotomous, meaning that when $\pst$ increases by $\gd$, $\pvt$ inversely decreases and vice versa. This finding provides evidence that large fluctuations in DST are governed by a simple dichotomy. We previously discovered another dichotomy for large fluctuations in frictionless particulate matter~\cite{rahbari_2017}. We showed that such dichotomies reveal unprecedented information about the collective behavior in rheological phase transitions, thus once again emphasizing that non-equilibrium phase transitions are described by simple dichotomies whose importance has been overlooked to date. We now focus on this novel dichotomy  to determine what lessons might be learned.\\

We start with the interpretation of a large fluctuation due to $\gd<0$. In a simple shear flow, each layer along the shearing direction has a larger drift velocity with respect to the layer beneath it, thus resulting in a positive local shear rate. However, when $\gd<0$, a given layer is slower than the one below it. Therefore, a large fluctuation of $\gd<0$ corresponds to a non-monotonic change in local drift velocity due to a retarded layer, thus resulting in a local negative power  $p_t<0$ providing $\delta\sxy >0$. In Fig.~\ref{fig:neg_power_t} for $\gd < \gd_c$,  $\pvt$ increases as $\gd_c$ is approached. This means that the flow becomes non-monotonic as the transition point is approached from below. This increasing non-monotonicity can be rationalized by the well-known instability near $\gd_c$~\cite{grob_2014, grob_2016, saw_2019}. 
In the solid-branch for $\gd>\gd_c$, the instabilities are washed out, and $\pvt$ decreases by increasing the shear rate. 
This finding is consistent with conventional wisdom, because the shear is a bias, and it removes all the retarded layers at large $\gd$,  thus explaining the decreasing trend in $\pvt$ in the solid branch.\\

%
%
%



Because of the dichotomy, $\pst$ has opposite behavior with respect to $\pvt$ across the transition region. To interpret the behavior of $\pst$, we first describe the relation of the couple-stress to the micro-physics. The couple-stress is related to the micro-physics via the total torque $\tau$ acting on particles according to
\begin{equation}
\sxyt - \sigma_{yx,t} \propto \sum_{i}\tau_{i},
\label{eq:coupled_stress}
\end{equation}
in which $i$ runs through all particles in the system. The relation between the couple-stress and torque has been well documented in the context of coarse-graining by Goldhirsch~\cite{goldhirsch_2010} and micropolar fluids by Mitarai, Hayakawa and Nakanishi~\cite{mitarai_2002}. Importantly, the left-hand side (lfh) of Eq.~\ref{eq:coupled_stress} is a coarse-grained field quantity, and the right-hand side is a particle property. Consequently, the results must be interpreted with caution. An alternative interpretation of the local shear rate is given by vorticity $\omega$ as 
\begin{equation}
\omega = \frac{\partial u_x}{\partial y} - \frac{\partial u_y}{\partial x},
\label{eq:virticity}
\end{equation}
where $u_x$ and $u_y$ are the drift velocity along the x- and y-directions, respectively. In a simple shear flow along the x-direction $\partial_x u_y= 0$; therefore, $\omega = \gd$. As a result, $ p_t \propto \tau\times \omega$ and $\pst$ correspond to a large fluctuation due to negative total torque of a sub-system whose vorticity is positive $\omega>0$. Re-inspection of Fig.~\ref{fig:neg_power_t} reveals that $\pst$, or equivalently $P( \tau^-, \omega^+)$, decreases as the transition point in the fluid branch is approached. This behavior is non-trivial, because it indicates that the instability enhances the uniformity of the rotational degrees of particles (it reduces negative events due to inverse torque). \\


%
%

\begin{figure}[ht]
	\centering
	\hfill\includegraphics[width=.5\textwidth]{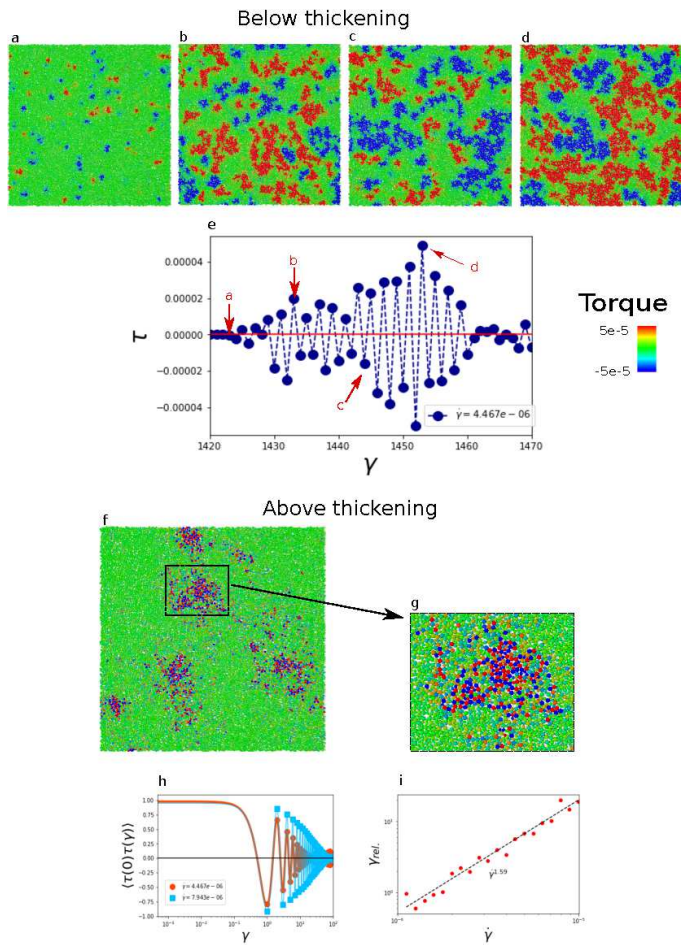}
	\hfill\\
	\caption[]{{\bf Torque across thickening} Panels-a to -d show subsequent snapshots of the system as it undergoes instability at  $\gd = 4.467\times 10^{-6}$. The color coding corresponds to the  total torque of each particle. As the system is sheared, domains of like-torque particles nucleate and grow, thus resulting in an enhancement of the rotational degrees of freedom, which we show to underlie the well-known instability near the thickening transition. The mean torque as a function of strain is shown in panel-e.  In panel-f, a snapshot of the system is shown after thickening for $\gd = 1.122 \times 10^{-5}$. The system is homogeneous except for localized clusters of opposite-torque particles. A closeup view of such a cluster is given in panel-g, which shows that the instability diminishes in the solid branch. The auto-correlation function of the total torque of the system shows a damped-oscillatory behavior in the fluid phase (panel-h) whose relaxation strain scales with $\gd$ with an exponent of $1.59$.
		\label{fig:snapshot1}}
\end{figure}

To gain a better understanding of the enhancement of the uniformity of rotational degrees of freedom, we display subsequent snapshots as the system is sheared from left to right  in Fig.~\ref{fig:snapshot1}-a to -d. The color coding corresponds to the total torque of each particle in which the blue particles have $\tau \leq -5\times 10^{-5}$, the red particles have  $\tau \geq 5\times 10^{-5}$, and the green particles have nearly zero torque. The shear rate is $\gd = 4.467\times 10^{-6}$, which is below $\gd_c$. In snapshot-a, the system is homogeneous except for anomalies appearing as tiny clusters of large negative and positive like-torque particles. These domains of large torque particles show that large torque is spatially localized. These clusters grow spatially in snapshot-b, with more red (positive) clusters. In contrast, in snapshot-c, the blue (negative) clusters appear, and finally in snapshot-d, the red clusters nearly percolate in the system. These clusters of like-torque particles are analogous to domains of like-spin sites in an Ising model at finite temperature. To show how the total torque of the system changes in these snapshots, we display the mean torque in the system as a function of strain in panel-e. In this figure, the corresponding mean torque of the snapshots is marked by red letters/arrows.  At $\tau = 1420$, the mean torque is zero; however, it begins to exhibit an oscillatory behavior whose amplitude is first enhanced, then  decays and finally fades to zero at $\gamma=1455$. Moreover, the torque subsequently undergoes another oscillatory behavior for $\gd >1470$. This is a typical pattern that repeats throughout the simulations for $\gd < \gd_c$. Indeed, we see that the mean torque is positive at snapshot-b in which the red (positive) clusters predominate the system; in contrast, in snapshot-c, where the blue (negative) clusters predominate the system, the mean torque is negative. The mean torque reaches its maximum in snapshot-d, where the red (positive) clusters nearly percolate in the system. Here, we reach an important conclusion: the oscillatory behavior of the torque, which is reminiscent of  rheochaos~\cite{cates_2002}, originates from the collective behavior of clusters of like-torque particles. Moreover, from a critical phenomena viewpoint, these spatially extended clusters resemble  domains of like-spin sites in the Ising model at finite temperature. \\


In snapshot-f, a typical configuration of the system is displayed after thickening. The shear rate is $\gd = 1.122 \times 10^{-5}$. In this snapshot, the system is homogeneous except for a few anomalous domains of opposite-torque particles. Interestingly, in these clusters, particles with anomalously large positive torque co-exist with those of anomalously large negative torque. A closeup view of such a cluster is given in panel-g. These regions of very large positive and negative torque particles may be compared to the Ising model at infinite temperature. Now we are ready to explain the behavior of the effective temperature $T_e$ in Fig.~\ref{fig:tot_power_FT}-b. As we discussed earlier, the effective temperature increases more than two orders of magnitude, and this increase might imply that the thickening  transition is equivalent to an equilibrium phase transition from a finite temperature to the infinite temperature--- an order-disorder-type transition. Indeed, this argument is supported by the snapshots across the thickening that imply a transition from a state of like-torque particles to that of opposite-torque particles. To characterize the oscillation of torque, we compute the auto-correlation function of torque $\left< \tau(0) \tau(\gamma)\right>$ in panel-h for two shear rates below thickening, 
which, as might be expected, is a damped oscillation of the form 
\begin{equation}
\left< \tau(0) \tau(\gamma)\right> = e^{-\gamma/\gamma_{rel.}} \cos(\pi\gamma),
\label{eq:auto_torque}
\end{equation}
where $\gamma_{rel.}$ is a relaxation strain. Solid symbols indicate our data from simulations, and solid lines correspond to fits via Eq.~\ref{eq:auto_torque}. The $\gamma_{rel.}$ increases as the transition point at $\gd = \gd_c$ is approached. In panel-i $\gamma_{rel.}\propto \gd^{1.59}$ for $\gd<\gd_c$ in the fluid phase. This damped oscillation is a direct consequence of the instability, and its increasing relation with $\gd$ is consistent with the findings from recent experiments reporting increased  instability as the transition point is approached~\cite{rathee_2017, saint_2018}. In the solid phase, the auto-correlation function becomes a step function reminiscent of a Markov-process, thus indicating that the instability vanishes in the solid-phase after thickening. \\

%

%
%

We now demonstrate the interplay between the above-mentioned collective behavior of the rotational degrees of freedom of particles and the rheology of the system. In Fig.~\ref{fig:snapshot2}, we display snapshots of the system with color coding corresponding to the total torque (first row) and shear stress (second row) of each particle . We display the shear stress per particle, which is different from the coarse-grained shear stress over a domain. Panels-a to -d row $1$ show the total torque per particle as the system enters the instability region, where the total torque in the system oscillates ( according to Fig.~\ref{fig:snapshot1}-e). Each snapshot shows a configuration of the system after $\delta \gamma = 1$ for $\gd = 4.467\times 10^{-6}$. Remarkably, in panel-a row $2$, where the same configuration is colored by the total shear stress per particle, particles with large shear stress form stress-bearing chains along the compression direction (yellow chains). In snapshot-b, clusters of like-torque particles become larger, and as a result, the stress-bearing chains become more pronounced, with a color shift to larger stresses (red) along the compression direction. Notably, as a result of larger negative like-torque clusters in panel-b, negative stress-bearing chains in blue (negative stress) form along the dilation direction. These red and blue stress-bearing structures become more pronounced in panels-c and -d, where clusters of like-torque particles become larger. In addition, the configuration of the system changes dramatically from one snapshot to the other below thickening. In panels-e to -h, we display similar snapshots after thickening for  $\gd = 1.122 \times 10^{-5}$. Each snapshot shows a configuration of the system after a strain difference of $\delta \gamma = 2$. There are strong stress-bearing structures along the compression direction (red chains), and particles with negative stress also form chains along the dilation direction (blue chains). Moreover, chains of positive stress percolate through the system, whereas chains of negative stress do not. Interestingly, even though the strain difference between each snapshot here is twice that in panels-a to -d, the structure of stress-bearing chains does not change dramatically from one snapshot to the next,  because  the system is in solid state. To quantitatively determine the correspondence between like-torque clusters and stress-bearing structures in the fluid phase, we plot the second moment of the torque  $\left<\sxy^2\right>$ and stress  $\left<\tau^2\right>$ per particle versus the strain
in Supplementary Fig.~4. It can be seen that $\left<\sxy^2\right>$ and $\left<\tau^2\right>$ change proportionately  as a function of the strain, thus indicating that fluctuations in torque directly influence the rheology, and the oscillation of torque results in spatiotemporal fluctuations in stress. However, in the solid phase, Supplementary Fig.~5, these quantities are independent of one another, and we do not observe an appreciable correlation between the torque and stress.\\

{\bf Conclusion.} We investigated fluctuations of power in a model
system undergoing DST. We showed that large fluctuations caused by
frictional forces are governed by a simple dichotomy that underlies
the novel collective behaviors across the thickening transition. The
joint probability of large fluctuations due to frictional forces
$\pst$ decreases as the thickening transition is approached, thus
indicating an enhanced uniformity of rotational degrees of
freedom. Consequently, we discovered clusters of like-torque particles
akin to an Ising model in finite temperature. We showed that (1) the
growth of these clusters directly correlates with the rheology (2) the
formation of the clusters results in spatially heterogeneous
structures of stress-bearing chains below the thickening and (3) a
competition between the opposite like-torque clusters underlies the
origin of temporal fluctuations. Accordingly, we have identified the
origin of spatiotemporal fluctuations near the thickening transition.
After thickening in the solid state, we observe clusters of
opposite-torque particles, analogously to an Ising model at infinite
temperature. Moreover, in this regime, particles grind against each
other, as they would be obliged to do by frictional forces that glue
particles to one another. This opposite motion is similar to that of
{\em gear wheels} in a mechanical watch. The increase of the effective
temperature by more than two orders of magnitude at the transition
point can thus be explained as the system transitions from the fluid
state, in which large clumps of particles rotate rigidly in one or the
other direction, to a solid state, in which particles grind against
each other. \\




We now conclude with some points discussing the possible implications of our results for a broader audience:\\

\begin{enumerate}[(i)]

\item{ {\bf Order-disorder scenario.} Besides the Wyart-Cates picture, several scenarios have been proposed  to describe  the  mechanisms underlying shear thickening. Two well-known mechanisms are (1) cluster formation due to hydrodynamic interactions~\cite{cheng_2011} and (2) an order-disorder transition~\cite{hoffman_1974, hoffman_1998}. In these scenarios, particles form ordered structures at small flow velocities, and these structures become unstable at large flow velocities. Although the formation of such ordered structures cannot occur in a bidisperse system, the similarity of these scenarios to our proposed order-disorder transition is notable. However, our order-disorder scenario involves only a transition in rotational degrees of freedom, in contrast to the above-mentioned  mechanisms  involving steric preferences of translational degrees of freedom.\\ 
}

\begin{figure}[ht] 
	\centering
	\hfill\includegraphics[width=.5\textwidth]{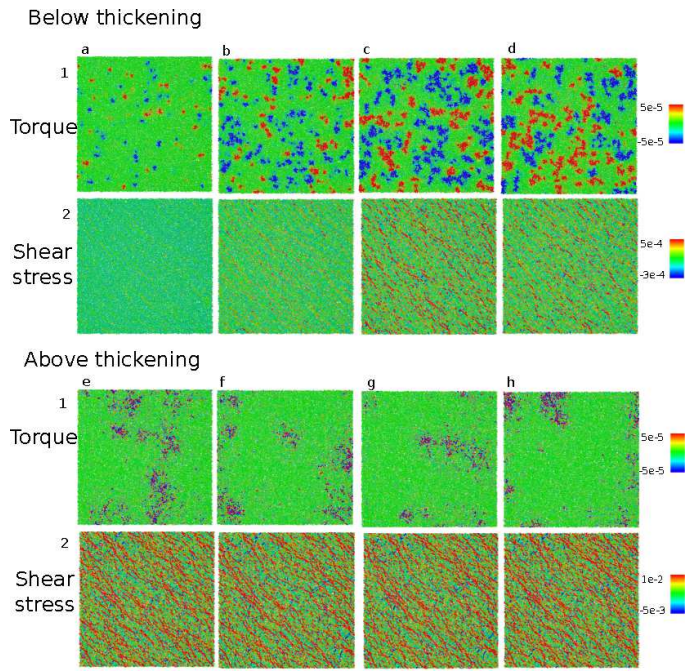}
	\hfill\\
	\caption[]{{\bf Torque-stress snapshots .} Snapshots of the system both below and above thickening. The color code in row number $1$ in each set corresponds to the total torque of each particle, and row number $2$ corresponds to the total shear stress per particle. In panels-a to -d, the shear rate is $\gd = 4.467\times 10^{-6}$, and the difference between each snapshot from left to right is a strain difference of unit length $\delta\gamma = 1$. The total torque of the system is given in panel-e of Fig.~\ref{fig:snapshot1}, which corresponds to the nucleation  of the instability. Clusters of like-torque grow larger as the instability sets in. In panel-b, larger clusters than those in panel-a result in stress-bearing structures. Remarkably, very large positive stress-bearing chains form along the compression direction (red chains), and shorter chains of very large negative stress form along the dilation direction (blue chains). In panel-c, the clusters of very large negative like-torque particles have become larger, thus resulting in an appreciable enhancement in very large negative stress-bearing structures along the dilation direction (blue chains). The second set at the bottom shows similar snapshots for $\gd = 1.122 \times 10^{-5}$ above thickening, and the difference between each snapshot from left to right is  $\delta\gamma = 2$. Locally, like-torque clusters are absent, and instead we see domains of particles with very large positive and negative torques. In contrast to the stress-bearing chains below thickening, those structures here seem to have no correlation with the torque, and their configurations persist for a long time. 
	\label{fig:snapshot2}}
\end{figure}



%
%
%
%
%
%

\item{  {\bf Yielding of frictional systems.} We note that Chattoraj \etal have recently demonstrated that a deformed very dense frictional system exhibits oscillatory instability near yielding, as a result of a pair of complex eigenvalues of the Hessain matrix~\cite{chattoraj_2019, chattoraj_2019_2}; this is in contrast to a frictionless system whose Hessian has only real eigenvalues, and a failure occurs when one of those eigenvalues becomes zero--- a saddle node bifurcation~\cite{maloney_2004}. Chattoraj \etal  
have also discussed a possible relation between the oscillatory amplifications in a frictional system with a long-standing problem in earthquake physics, remote triggering~\cite{felzer_2006}. We expect that our discovery of the collective behavior of the rotational degrees of freedom of frictional particles might also shed light on the micro-physics of the oscillatory instability near the yielding transition.

}


\item{  {\bf Order parameter of non-equilibrium.} Collective behaviors in equilibrium phase transitions are described by order parameters~\cite{kardar_2007}. Non-equilibrium phase transitions have a much richer phenomenology, and as  a result of that their collective behavior, cannot be explored by a single order parameter. Instead, as we demonstrated here, distributions of some order-like parameters must be examined. We showed that  such examination can be performed well by using the stochastic thermodynamics of some stochastic energetics parameters. The investigation of large fluctuations in such  stochastic energetics parameters, such as power, led us to the discovery of dichotomies describing the underlying collective modes. The scheme developed here can be easily applied to a much wider class of non-equilibrium phase transitions. \\

}

\item{  {\bf Effective temperatures.} The 21st century is the era of non-equilibrium statistical physics, which has a  wide focus on many real-life, everyday phenomena. Yet, equilibrium thermodynamics is key to understanding the nature of non-equilibrium phenomena. A quantitative means of establishing such a connection is provided by the so-called effective temperatures that bridge equilibrium and non-equilibrium worlds~\cite{makse_2002}. The first  such attempt may be the work by Edwards and Oakeshott in
an acclaimed paper that proposed that the principles of equilibrium statistical mechanics can be applied to granular materials~\cite{edwards_1989}. The authors suggested that the packing
fraction in granular materials may play the same role as energy in equilibrium statistical mechanics and defined the analogous density of states $\Omega(\phi) = \sum_{\nu} \delta(\phi - \phi_v)$, where the subscript $\nu$ corresponds to jammed states with the condition of force and torque balance being satisfied~\cite{bi_2015_2}. Edwards' entropy can then be defined as $S = \ln \Omega(\phi)$, which results in a temperature-like quantity $X$ via $1/X = \partial
S(\phi) / \partial \phi$. Later, because of the importance of normal and tangential forces in configurations of jammed states, stress was considered as an additional state variable. The resulting
temperature, known as angoricity, corresponds to a jammed canonical ensemble.  Several methods have been developed to test the ideas of Edwards. Probably the most renowned method is the overlapping histogram method by Dean and Leferevre~\cite{dean_2003}, which has been used in many recent studies~\cite{zhao_2012, mcnamara_2009, bililign_2019}. Although Edwards' original intention was to apply ensemble theory to describe the dynamics of slowly driven granular materials, in which the system moves from one jammed state into
another, most recent literature has focused only on static jammed
states. Furthermore, some recent work has suggested that angoricity is analogous to an effective temperature that describes the strength of the mechanical noise in driven granular
materials~\cite{behringer_2008, zheng_2018_2}. However, this approach is in early stages of development. One large obstacle to applying
Edwards' theory is that the density of states $\Omega(\phi)$ and the corresponding partition function may be unknown for a given granular ensemble~\cite{bi_2015_2}. Our stochastic thermodynamic approach provides an alternative way to compute an effective stress-temperature. We showed here and in a previous report~\cite{rahbari_2017} that the resulting effective temperature is stress-like, \ie, it remains constant at the vanishing limit of the shear rate in the jammed configurations. However, it is Bagnoldian in the fluid phase. This method provides an alternative approach to investigate fluctuations in a wide range of driven non-equilibrium systems. The main advantage of this approach is that it does not require strictly jammed states.\\
}

\end{enumerate}






{\bf Acknowledgments} We acknowledge fruitful discussions with Hyunggyu Park, Mike Cates,  Hisao Hayakawa, Ludovic Berthier, Michio Otsuki, Takeshi Kawasaki, Abbas Ali Saberi, and Ji Woong Yu.  

%


\input{DST_habib_v3_edited_letter_form.bbl}



\pagebreak
\clearpage

\begin{center}
	\textbf{\large Supplemental Materials: Origin of spatiotemporal fluctuations in Discontinuous Shear Thickening}
\end{center}


\setcounter{equation}{0}
\setcounter{figure}{0}
\setcounter{table}{0}
\setcounter{page}{1}
\makeatletter
\renewcommand{\theequation}{S\arabic{equation}}
\renewcommand{\thefigure}{S\arabic{figure}}
\renewcommand{\bibnumfmt}[1]{[S#1]}
\renewcommand{\citenumfont}[1]{S#1}


\section{Methods}
\label{sec:methods}

We use a linear dashpot-spring to model both normal and tangential forces~\cite{poeschel2005}. Two particles at positions ${\bf r_i}$  and ${\bf r_j}$ with radii $a_i$ and $a_j$, respectively, interact when they overlap,  $\delta = |{\bf r_i} - {\bf r_j} | - (a_i + a_j) <0 $. A spring whose force is proportional to the overlap $\delta$ acts as a repulsive mechanism between two colliding particles. The interaction force along the normal direction is then given by
\begin{equation}
f_{ij,n} = k_n \delta - \eta_n ({\bf v_i} - {\bf v_j}) \cdot {\bf n}_{ij},
\end{equation}
where $k_n$ and $\eta_n$ are the elastic, and damping constants, respectively, and ${\bf n}_{ij}$ is a unit vector along the line connecting the centers of two particles ${\bf n}_{ij} = ({\bf r_i} - {\bf r_j}) / |{\bf r_i} - {\bf r_j} |$.

With ${\bf \omega_i}$ and ${\bf \omega_j}$, the angular velocities, the total tangential velocity at the contact point can be written as 
\begin{equation}
{\bf v}_{ij,t} = ({\bf I} - {\bf n}_{ij}{\bf n}_{ij} ) \cdot[  {\bf v_i - v_j} - ( a_i{\bf \omega_i} + a_j{\bf \omega_j}) \times {\bf n}_{ij} ].
\end{equation}

Integrating the tangential velocity ${\bf v}_{ij,t}$ from the initiation of contact to the current time gives the tangential overlap as $\xi = \int_{0}^{t_{coll}} | {\bf v}_{ij,t}| dt^\prime$. A spring proportional to $\xi$ acts in the tangential direction along the contact plane to model the static friction
\begin{equation}
f_{ij,t} = k_t \xi - \eta_t {\bf v}_{ij,t}\cdot {\bf t}_{ij},
\end{equation} 
where $k_t$ and $\eta_t$ are the spring and damping coefficients, and ${\bf t}_{ij}$ is a unit vector along the contact plane, ${\bf t}_{ij} \cdot {\bf n}_{ij} = 0$. A torque proportional to the tangential force acts on each particle. Accordingly, the total force is equal to
\begin{equation}
{\bf f}_{ij} = f_{ij,n} {\bf n}_{ij} +  f_{ij,t} {\bf t}_{ij},
\end{equation}
from which translational and rotational degrees of freedom are coupled in this model.\\

We use a $50:50$ bidisperse mixture of particles whose ratio of radii is $1.4$. The diameter of small particles is chosen to be the unit of length. The mass is equal to the area of each particle. The spring constants are $k_n = 1$ and $k_t = 0.5k_n$ and the damping coefficients are $\eta_n = \eta_t = 1$. The magnitude of the tangential force is bound by Coulomb's frictional law $|f_{ij,t}|\le \mu |f_{ij,n}|$ where $\mu = 1$ is the friction coefficient.

We use Large-scale Atomic/Molecular Massively Parallel Simulator (LAMMPS) to integrate equations of motion of particles. This process is done by using pair style  $gran/hooke/history$ to model the interactions plus Lees-Edwards boundary conditions by using $deform$.

\section{Complementary results}

\begin{figure}[ht]
	\centering
	\hfill\includegraphics[width=0.475\textwidth]{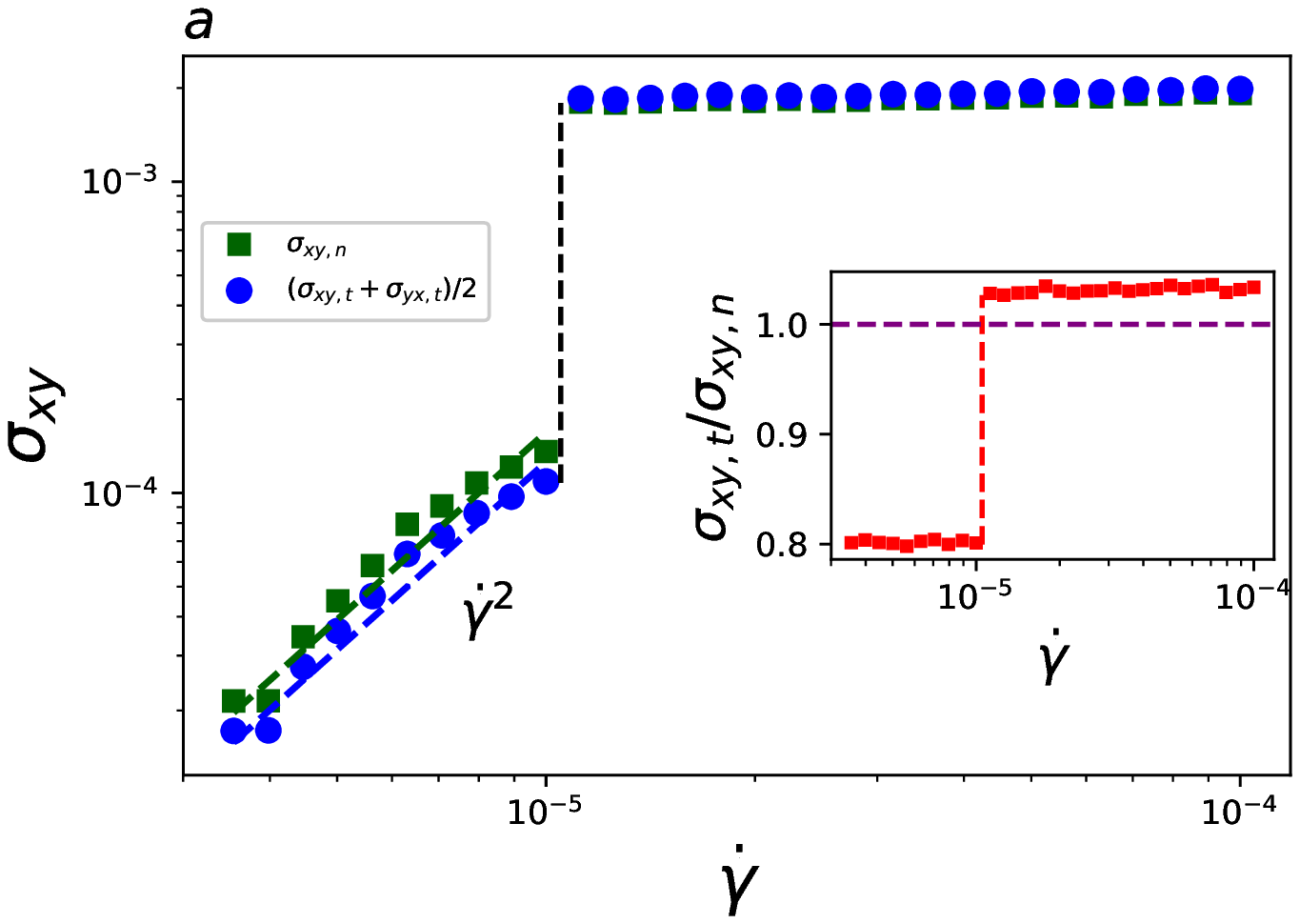}
	\hfill\includegraphics[width=0.475\textwidth]{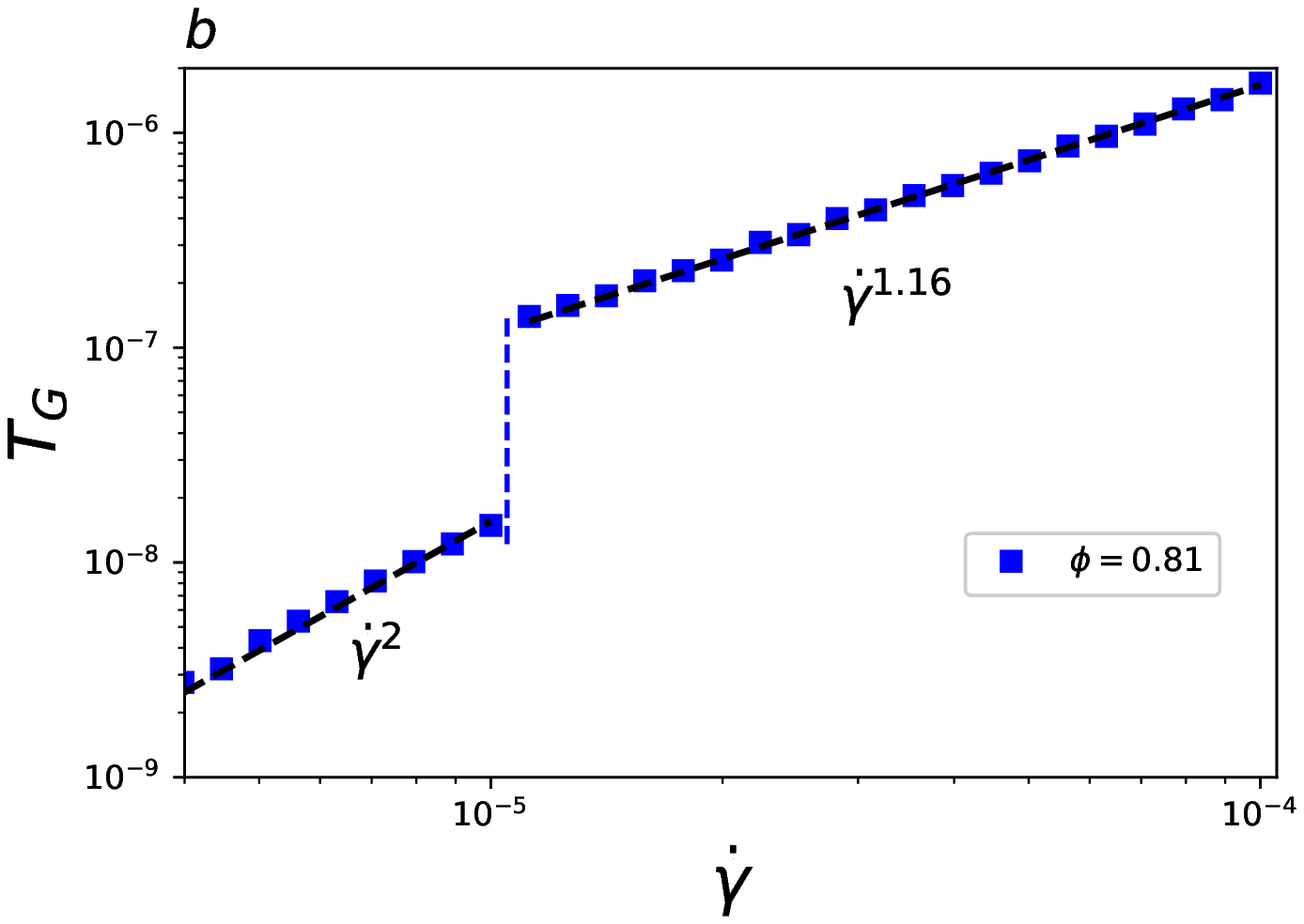}
	\hfill\\
	\caption[]{{\bf  Rheology across thickening.} (a) Shear stress as a function of the shear rate for normal (squares) and tangential (circles) components. For $\gd < \gd_c$, the behavior is Bagnoldian ($\gd^2$): a typical fluid-like behavior. For $\gd > \gd_c$, the system is solid-like and exhibits a finite yield stress. The inset shows that below thickening, the normal component is dominant, whereas above the transition, the tangential part dominates. (b) Kinetic temperature as a function of shear rate. Whereas for $\gd < \gd_c$, the behavior is Bagnoldian, in agreement with the rheology, above $\gd_c$, it shows a dense fluid behavior not consistent with the rheology. The packing fraction is $\phi = 0.81$, and the number of particles is $N = 16384$. 
		\label{fig:mean_Sxy}}
\end{figure}

\begin{figure}[ht]
	\centering
	\hfill\includegraphics[width=0.45\textwidth]{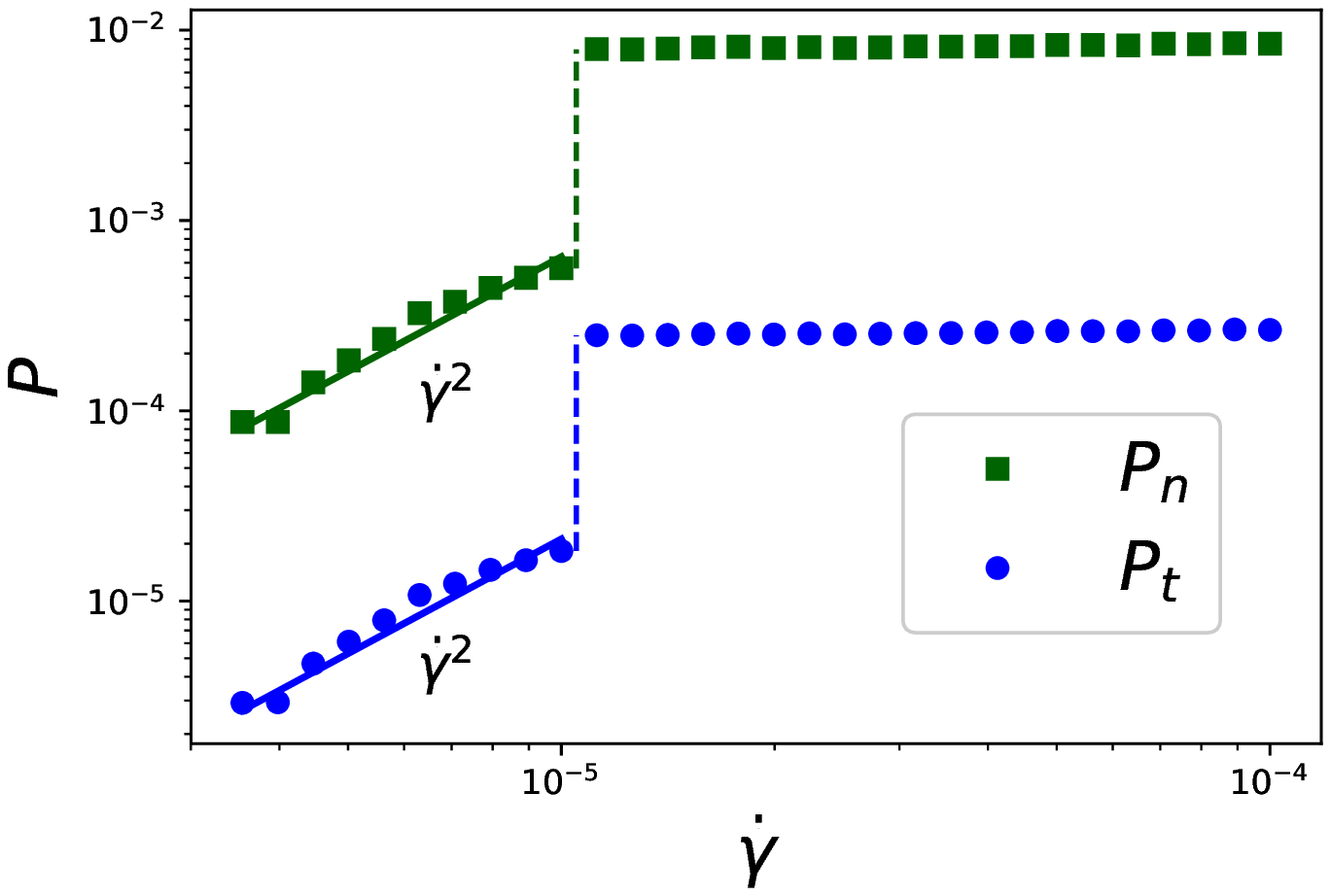}
	\hfill\\
	\caption[]{{\bf Rheology across thickening.}  Pressure defined as $(\sigma_{xx} + \sigma_{yy} )/2$ for both normal and tangential components given by circles and squares, respectively. Below thickening, Bagnold behavior ($\gd^2$) is seen; above the transition, the pressure shows a behavior of Herschel-Bulkley form. In contrast to the shear stress, the normal part of the pressure is dominant both above and below the transition. 
		\label{fig:P_t_n}}
\end{figure}

A typical flow curve of the system is given in Fig.~\ref{fig:mean_Sxy}-a. The simulations start at a relatively high shear rate $\gd = 10^{-4}$, and the shear rate is decreased stepwise such that the steps are equidistant on a logarithmic scale. Each simulation runs for a total strain of $\gamma_{tot}=20L$, and we store a snapshot of the system at each strain increment equal to the unit of length $\delta\gamma = 1$. There exists a critical shear rate $\gd_c \simeq 10^{-5}$ at which the stress abruptly changes by more than one order of magnitude. This finding  is consistent with those reported by Otsuki and Hayakawa~\cite{otsuki_2011}. The packing fraction is $\phi=0.81$ and the system shows DST for  densities above a fictitious critical density of $\phi_c=0.795$, again in agreement with~\cite{otsuki_2011}. The critical shear rate separates two distinct fluid and solid states. For $\gd < \gd_c$, a Bagnoldian rheology $\sigma_{xy}, P \propto \gd^2$ is observed, thus indicating a fluid-like behavior. For $\gd > \gd_c$, the system is solid-like, and the flow curves are generally described by  Herschel-Bulkley rheology $\sigma_{xy}, P = \sigma_y + k\gd^y$ with a distinct offset $\sigma_y$ that indicates  the appearance of shear-driven yield stress. We separately measure the contributions of tangential and repulsive interactions to the  stress tensor. Green squares and blue circles correspond to repulsive, $\sxyn$, and tangential,  $(\sxyt + \syxt)/2$, shear stress, respectively,  and the inset shows the ratio of the two. For $\gd<\gd_c$, the normal component is dominant; however, above $\gd_c$, the tangential component  predominates in the momentum transfer. This finding is in contrast to the pressure in which the normal component is dominant both above and below the thickening (Appendix Fig.~\ref{fig:P_t_n}). These results once again emphasize the canonical role played by friction in DST. \\

Jamming of frictional disks occurs at a coordination number of $z_J = d+1 = 3$~\cite{vanhecke_2010}. Across the transition region in DST, the coordination number also increases from $z = z_f(\gd)<z_{J}$ in the fluid state to $z = z_s(\gd)>z_{J}$ in the solid branch (see Appendix Fig.~\ref{fig:coordination}), thus demonstrating that 	shear thickening is a consequence of a shear driven jamming transition~\cite{bi_2011}, which is generally accompanied by proliferation of the contacts~\cite{seto_2013}. All aforementioned physical measures faithfully present the inherent state of the system as it transitions from the fluid to the solid states. However, the kinetic temperature of the system in Fig.~\ref{fig:mean_Sxy}-b, known as the granular temperature $T_G$, has misleading behavior for $\gd> \gd_c$. Whereas it shows a quadratic dependence below $\gd_c$, consistently with the rheology, above $\gd_c$, $T_G$ has an algebraic dependence on $\gd$ with an exponent of $1.16$, which is reminiscent of the behavior of a dense fluid. This finding is in contrast to the jammed nature of the system above $\gd_c$. Therefore, the kinetic temperature cannot be used as a real measure of fluctuations in this system, and an alternative measure is required. To resolve this discrepancy, an effective temperature that complies with the rheology is required. To this end, we investigate large fluctuations in injected power $p$ by using the tools of stochastic thermodynamics to derive the effective temperature. \\

In this appendix, we display some auxiliary figures to complement  the descriptions of the main figures in this manuscript. 

\begin{figure}[ht]
	\centering
	\hfill\includegraphics[width=0.45\textwidth]{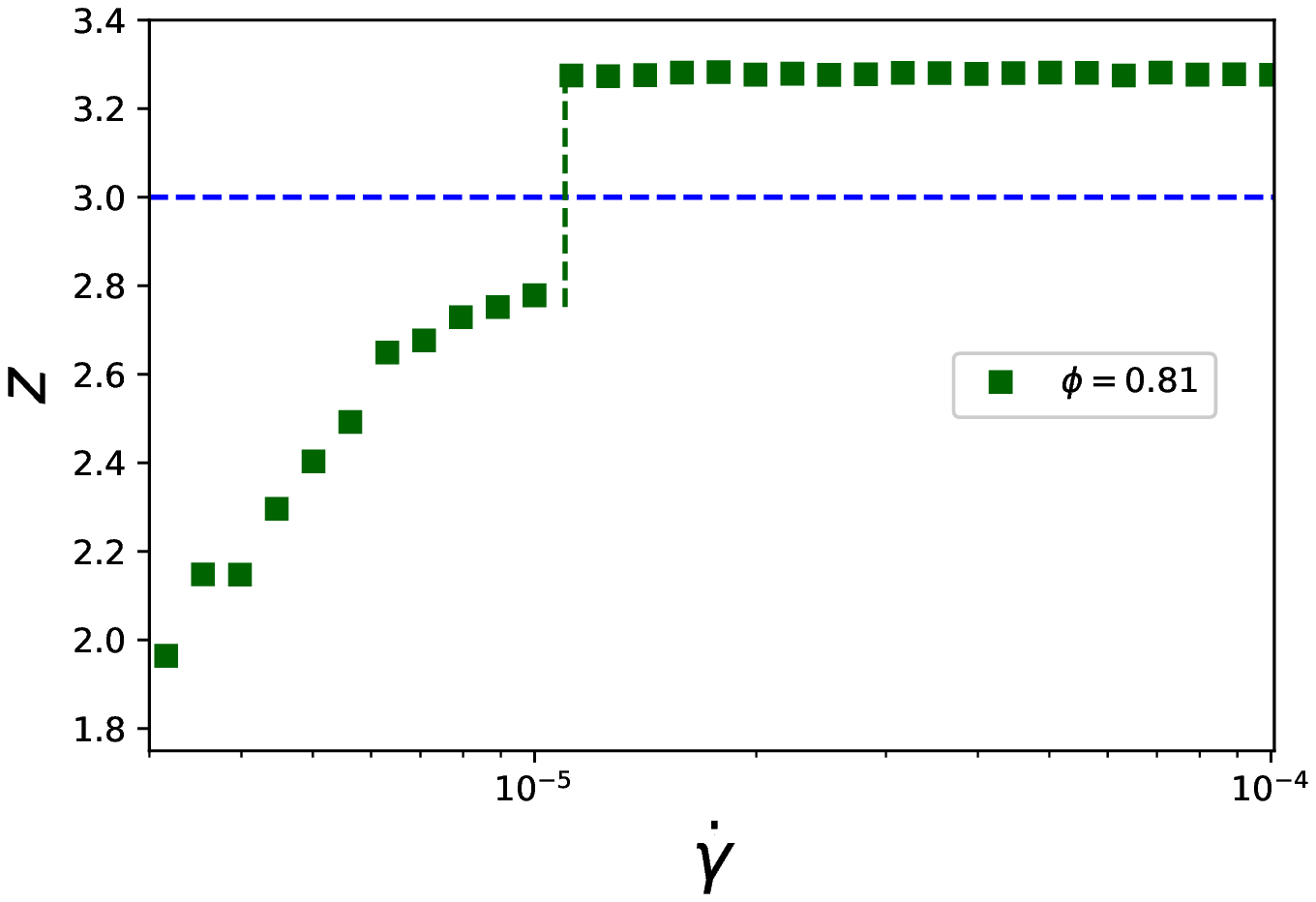}
	\hfill\\
	\caption[]{{\bf Coordination number.} The mean coordination numbers as a function of the shear rate. The jamming coordination number for this system is $z_J = 3$. Below thickening, $z<z_J$. At the transition point, the coordination number increases to $z>z_J$. 
		\label{fig:coordination}}
\end{figure}

\begin{figure}[ht] 
	\centering
	\hfill\includegraphics[width=0.5\textwidth]{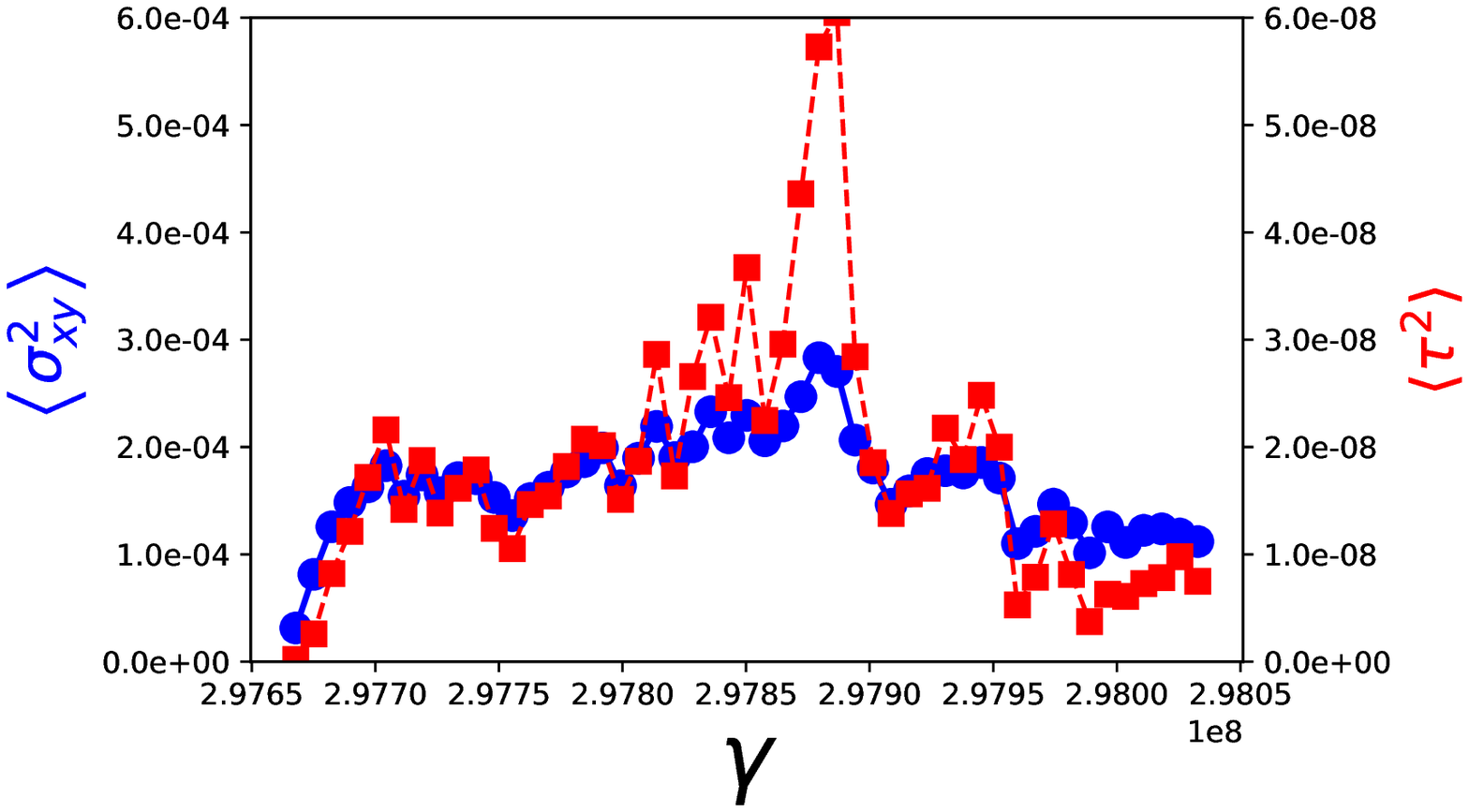}
	\hfill\\
	\caption[]{{\bf Correlated behavior of stress and torque in the fluid phase.} To quantitatively show that the formation of like-torque clusters results in stress heterogeneities, we display the second-moment of shear stress per particle (circles) and total torque of each particle (squares). These two quantities are inter-correlated, and any fluctuation in the total torque of particles accordingly results in that in the shear stress per particle. This figure is a quantitative demonstration of the correlation of stress and torque snapshots in Fig.~\ref{fig:snapshot2}.  The shear rate is $\gd = 4.467\times 10^{-6}$.
		\label{fig:mean_Sxy_per_part_fluid}}
\end{figure}

\begin{figure}[H] 
	\centering
	\hfill\includegraphics[width=0.5\textwidth]{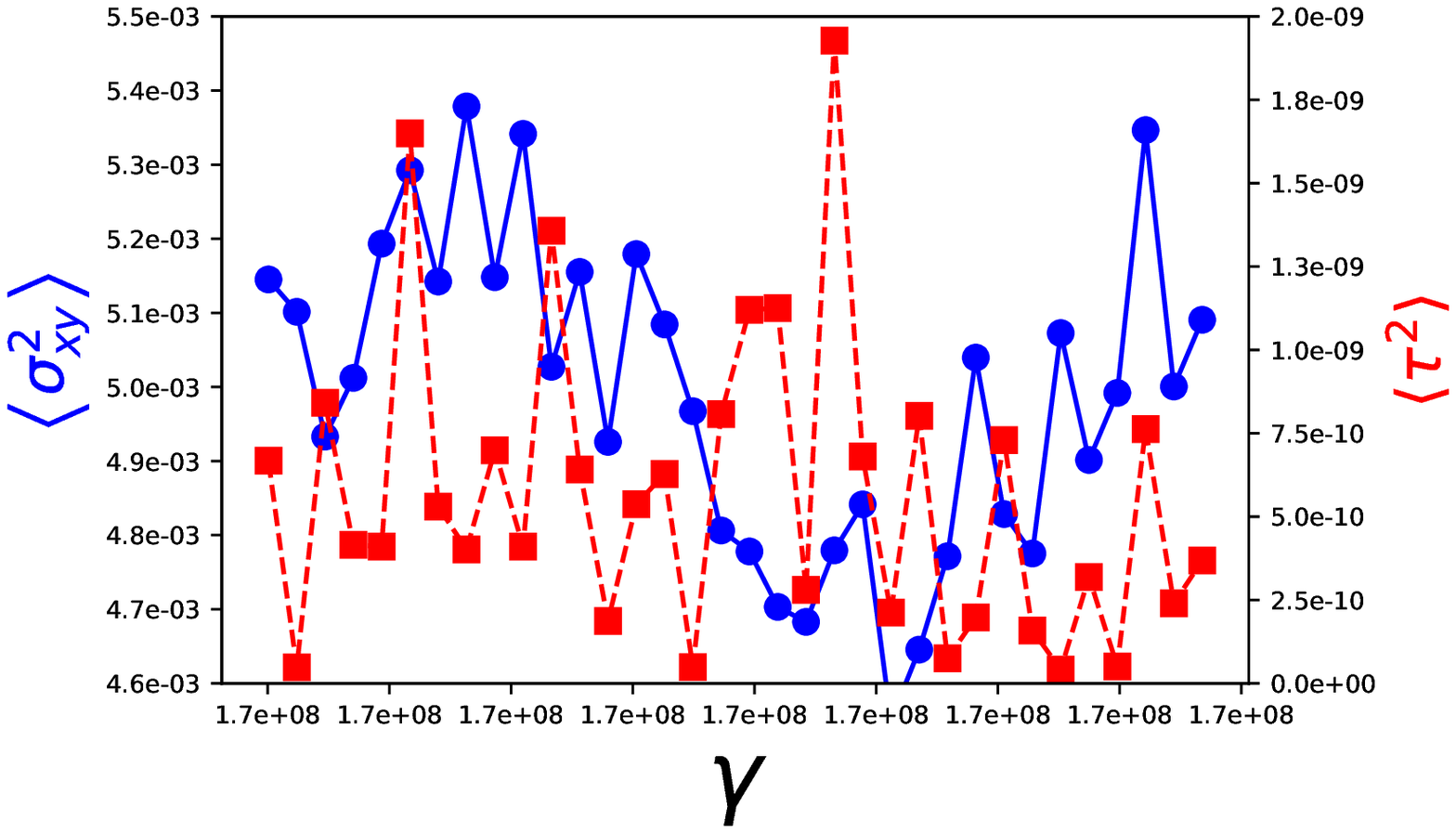}
	\hfill\\
	\caption[]{{\bf Un-correlated behavior of stress and torque in the solid phase.} Similarly to the previous figure, we display the second-moment of shear stress per particle (circles) and total torque of each particle (squares). The shear rate is $\gd = 1.122 \times 10^{-5}$ above thickening. Fluctuations in the shear stress per particle and torque can be seen to be un-correlated.
		\label{fig:mean_Sxy_per_part_solid}}
\end{figure}






\end{document}

%% file: DST_habib_v3_edited_letter_form.bbl
%

%% file: DST_habib_v3_edited_letter_form.bbl
\begin{thebibliography}{44}%
\makeatletter
\providecommand \@ifxundefined [1]{%
 \@ifx{#1\undefined}
}%
\providecommand \@ifnum [1]{%
 \ifnum #1\expandafter \@firstoftwo
 \else \expandafter \@secondoftwo
 \fi
}%
\providecommand \@ifx [1]{%
 \ifx #1\expandafter \@firstoftwo
 \else \expandafter \@secondoftwo
 \fi
}%
\providecommand \natexlab [1]{#1}%
\providecommand \enquote  [1]{``#1''}%
\providecommand \bibnamefont  [1]{#1}%
\providecommand \bibfnamefont [1]{#1}%
\providecommand \citenamefont [1]{#1}%
\providecommand \href@noop [0]{\@secondoftwo}%
\providecommand \href [0]{\begingroup \@sanitize@url \@href}%
\providecommand \@href[1]{\@@startlink{#1}\@@href}%
\providecommand \@@href[1]{\endgroup#1\@@endlink}%
\providecommand \@sanitize@url [0]{\catcode `\\12\catcode `\$12\catcode
  `\&12\catcode `\#12\catcode `\^12\catcode `\_12\catcode `\%12\relax}%
\providecommand \@@startlink[1]{}%
\providecommand \@@endlink[0]{}%
\providecommand \url  [0]{\begingroup\@sanitize@url \@url }%
\providecommand \@url [1]{\endgroup\@href {#1}{\urlprefix }}%
\providecommand \urlprefix  [0]{URL }%
\providecommand \Eprint [0]{\href }%
\providecommand \doibase [0]{http://dx.doi.org/}%
\providecommand \selectlanguage [0]{\@gobble}%
\providecommand \bibinfo  [0]{\@secondoftwo}%
\providecommand \bibfield  [0]{\@secondoftwo}%
\providecommand \translation [1]{[#1]}%
\providecommand \BibitemOpen [0]{}%
\providecommand \bibitemStop [0]{}%
\providecommand \bibitemNoStop [0]{.\EOS\space}%
\providecommand \EOS [0]{\spacefactor3000\relax}%
\providecommand \BibitemShut  [1]{\csname bibitem#1\endcsname}%
\let\auto@bib@innerbib\@empty
\bibitem [{\citenamefont {Brown}\ and\ \citenamefont
  {Jaeger}(2014)}]{brown_2014}%
  \BibitemOpen
  \bibfield  {author} {\bibinfo {author} {\bibfnamefont {E.}~\bibnamefont
  {Brown}}\ and\ \bibinfo {author} {\bibfnamefont {H.~M.}\ \bibnamefont
  {Jaeger}},\ }\bibfield  {title} {\enquote {\bibinfo {title} {Shear thickening
  in concentrated suspensions: phenomenology, mechanisms and relations to
  jamming},}\ }\href@noop {} {\bibfield  {journal} {\bibinfo  {journal} {Rep.
  Prog. Phys.}\ }\textbf {\bibinfo {volume} {77}},\ \bibinfo {pages} {046602}
  (\bibinfo {year} {2014})}\BibitemShut {NoStop}%
\bibitem [{\citenamefont {Morris}(2009)}]{morris_2009}%
  \BibitemOpen
  \bibfield  {author} {\bibinfo {author} {\bibfnamefont {J.~F.}\ \bibnamefont
  {Morris}},\ }\bibfield  {title} {\enquote {\bibinfo {title} {A review of
  microstructure in concentrated suspensions and its implications for rheology
  and bulk flow},}\ }\href@noop {} {\bibfield  {journal} {\bibinfo  {journal}
  {Rheol. acta}\ }\textbf {\bibinfo {volume} {48}},\ \bibinfo {pages}
  {909--923} (\bibinfo {year} {2009})}\BibitemShut {NoStop}%
\bibitem [{\citenamefont {Seto}\ \emph {et~al.}(2013)\citenamefont {Seto},
  \citenamefont {Mari}, \citenamefont {Morris},\ and\ \citenamefont
  {Denn}}]{seto_2013}%
  \BibitemOpen
  \bibfield  {author} {\bibinfo {author} {\bibfnamefont {R.}~\bibnamefont
  {Seto}}, \bibinfo {author} {\bibfnamefont {R.}~\bibnamefont {Mari}}, \bibinfo
  {author} {\bibfnamefont {J.~F.}\ \bibnamefont {Morris}}, \ and\ \bibinfo
  {author} {\bibfnamefont {M.~M.}\ \bibnamefont {Denn}},\ }\bibfield  {title}
  {\enquote {\bibinfo {title} {Discontinuous shear thickening of frictional
  hard-sphere suspensions},}\ }\href@noop {} {\bibfield  {journal} {\bibinfo
  {journal} {Phys. Rev. Lett.}\ }\textbf {\bibinfo {volume} {111}},\ \bibinfo
  {pages} {218301} (\bibinfo {year} {2013})}\BibitemShut {NoStop}%
\bibitem [{\citenamefont {Wyart}\ and\ \citenamefont
  {Cates}(2014)}]{wyart_2014}%
  \BibitemOpen
  \bibfield  {author} {\bibinfo {author} {\bibfnamefont {M.}~\bibnamefont
  {Wyart}}\ and\ \bibinfo {author} {\bibfnamefont {M.~E.}\ \bibnamefont
  {Cates}},\ }\bibfield  {title} {\enquote {\bibinfo {title} {Discontinuous
  shear thickening without inertia in dense non-{B}rownian suspensions},}\
  }\href@noop {} {\bibfield  {journal} {\bibinfo  {journal} {Phys. Rev. Lett.}\
  }\textbf {\bibinfo {volume} {112}},\ \bibinfo {pages} {098302} (\bibinfo
  {year} {2014})}\BibitemShut {NoStop}%
\bibitem [{\citenamefont {Otsuki}\ and\ \citenamefont
  {Hayakawa}(2011)}]{otsuki_2011}%
  \BibitemOpen
  \bibfield  {author} {\bibinfo {author} {\bibfnamefont {M.}~\bibnamefont
  {Otsuki}}\ and\ \bibinfo {author} {\bibfnamefont {H.}~\bibnamefont
  {Hayakawa}},\ }\bibfield  {title} {\enquote {\bibinfo {title} {Critical
  scaling near jamming transition for frictional granular particles},}\
  }\href@noop {} {\bibfield  {journal} {\bibinfo  {journal} {Phys. Rev. E}\
  }\textbf {\bibinfo {volume} {83}},\ \bibinfo {pages} {051301} (\bibinfo
  {year} {2011})}\BibitemShut {NoStop}%
\bibitem [{\citenamefont {Otsuki}\ and\ \citenamefont
  {Hayakawa}(2018)}]{otsuki_2018}%
  \BibitemOpen
  \bibfield  {author} {\bibinfo {author} {\bibfnamefont {M.}~\bibnamefont
  {Otsuki}}\ and\ \bibinfo {author} {\bibfnamefont {H.}~\bibnamefont
  {Hayakawa}},\ }\bibfield  {title} {\enquote {\bibinfo {title} {Shear jamming,
  discontinuous shear thickening, and fragile state in dry granular materials
  under oscillatory shear},}\ }\href@noop {} {\bibfield  {journal} {\bibinfo
  {journal} {arXiv preprint arXiv:1810.03846}\ } (\bibinfo {year}
  {2018})}\BibitemShut {NoStop}%
\bibitem [{\citenamefont {Kawasaki}\ and\ \citenamefont
  {Berthier}(2018)}]{kawasaki_2018}%
  \BibitemOpen
  \bibfield  {author} {\bibinfo {author} {\bibfnamefont {T.}~\bibnamefont
  {Kawasaki}}\ and\ \bibinfo {author} {\bibfnamefont {L.}~\bibnamefont
  {Berthier}},\ }\bibfield  {title} {\enquote {\bibinfo {title} {Discontinuous
  shear thickening in brownian suspensions},}\ }\href@noop {} {\bibfield
  {journal} {\bibinfo  {journal} {Phys. Rev. E}\ }\textbf {\bibinfo {volume}
  {98}},\ \bibinfo {pages} {012609} (\bibinfo {year} {2018})}\BibitemShut
  {NoStop}%
\bibitem [{\citenamefont {Saint-Michel}\ \emph {et~al.}(2018)\citenamefont
  {Saint-Michel}, \citenamefont {Gibaud},\ and\ \citenamefont
  {Manneville}}]{saint_2018}%
  \BibitemOpen
  \bibfield  {author} {\bibinfo {author} {\bibfnamefont {B.}~\bibnamefont
  {Saint-Michel}}, \bibinfo {author} {\bibfnamefont {T.}~\bibnamefont
  {Gibaud}}, \ and\ \bibinfo {author} {\bibfnamefont {S.}~\bibnamefont
  {Manneville}},\ }\bibfield  {title} {\enquote {\bibinfo {title} {Uncovering
  instabilities in the spatiotemporal dynamics of a shear-thickening cornstarch
  suspension},}\ }\href@noop {} {\bibfield  {journal} {\bibinfo  {journal}
  {Phys. Rev. X}\ }\textbf {\bibinfo {volume} {8}},\ \bibinfo {pages} {031006}
  (\bibinfo {year} {2018})}\BibitemShut {NoStop}%
\bibitem [{\citenamefont {Cates}\ \emph {et~al.}(2002)\citenamefont {Cates},
  \citenamefont {Head},\ and\ \citenamefont {Ajdari}}]{cates_2002}%
  \BibitemOpen
  \bibfield  {author} {\bibinfo {author} {\bibfnamefont {M.~E.}\ \bibnamefont
  {Cates}}, \bibinfo {author} {\bibfnamefont {D.~A.}\ \bibnamefont {Head}}, \
  and\ \bibinfo {author} {\bibfnamefont {A.}~\bibnamefont {Ajdari}},\
  }\bibfield  {title} {\enquote {\bibinfo {title} {Rheological chaos in a
  scalar shear-thickening model},}\ }\href@noop {} {\bibfield  {journal}
  {\bibinfo  {journal} {Phys. Rev. E}\ }\textbf {\bibinfo {volume} {66}},\
  \bibinfo {pages} {025202} (\bibinfo {year} {2002})}\BibitemShut {NoStop}%
\bibitem [{\citenamefont {Olmsted}(2008)}]{olmsted_2008}%
  \BibitemOpen
  \bibfield  {author} {\bibinfo {author} {\bibfnamefont {P.~D.}\ \bibnamefont
  {Olmsted}},\ }\bibfield  {title} {\enquote {\bibinfo {title} {Perspectives on
  shear banding in complex fluids},}\ }\href@noop {} {\bibfield  {journal}
  {\bibinfo  {journal} {Rheol. Acta}\ }\textbf {\bibinfo {volume} {47}},\
  \bibinfo {pages} {283--300} (\bibinfo {year} {2008})}\BibitemShut {NoStop}%
\bibitem [{\citenamefont {Chacko}\ \emph {et~al.}(2018)\citenamefont {Chacko},
  \citenamefont {Mari}, \citenamefont {Cates},\ and\ \citenamefont
  {Fielding}}]{chacko_2018}%
  \BibitemOpen
  \bibfield  {author} {\bibinfo {author} {\bibfnamefont {R.~N.}\ \bibnamefont
  {Chacko}}, \bibinfo {author} {\bibfnamefont {R.}~\bibnamefont {Mari}},
  \bibinfo {author} {\bibfnamefont {M.~E.}\ \bibnamefont {Cates}}, \ and\
  \bibinfo {author} {\bibfnamefont {S.~M.}\ \bibnamefont {Fielding}},\
  }\bibfield  {title} {\enquote {\bibinfo {title} {Dynamic vorticity banding in
  discontinuously shear thickening suspensions},}\ }\href@noop {} {\bibfield
  {journal} {\bibinfo  {journal} {Phys. Rev. Lett.}\ }\textbf {\bibinfo
  {volume} {121}},\ \bibinfo {pages} {108003} (\bibinfo {year}
  {2018})}\BibitemShut {NoStop}%
\bibitem [{\citenamefont {Rathee}\ \emph {et~al.}(2017)\citenamefont {Rathee},
  \citenamefont {Blair},\ and\ \citenamefont {Urbach}}]{rathee_2017}%
  \BibitemOpen
  \bibfield  {author} {\bibinfo {author} {\bibfnamefont {V.}~\bibnamefont
  {Rathee}}, \bibinfo {author} {\bibfnamefont {D.~L.}\ \bibnamefont {Blair}}, \
  and\ \bibinfo {author} {\bibfnamefont {J.~S.}\ \bibnamefont {Urbach}},\
  }\bibfield  {title} {\enquote {\bibinfo {title} {Localized stress
  fluctuations drive shear thickening in dense suspensions},}\ }\href@noop {}
  {\bibfield  {journal} {\bibinfo  {journal} {Proc. Nat. Acad. Sc.}\ }\textbf
  {\bibinfo {volume} {114}},\ \bibinfo {pages} {8740--8745} (\bibinfo {year}
  {2017})}\BibitemShut {NoStop}%
\bibitem [{\citenamefont {Grob}\ \emph {et~al.}(2014)\citenamefont {Grob},
  \citenamefont {Heussinger},\ and\ \citenamefont {Zippelius}}]{grob_2014}%
  \BibitemOpen
  \bibfield  {author} {\bibinfo {author} {\bibfnamefont {M.}~\bibnamefont
  {Grob}}, \bibinfo {author} {\bibfnamefont {C.}~\bibnamefont {Heussinger}}, \
  and\ \bibinfo {author} {\bibfnamefont {A.}~\bibnamefont {Zippelius}},\
  }\bibfield  {title} {\enquote {\bibinfo {title} {Jamming of frictional
  particles: A nonequilibrium first-order phase transition},}\ }\href@noop {}
  {\bibfield  {journal} {\bibinfo  {journal} {Phys. Rev. E}\ }\textbf {\bibinfo
  {volume} {89}},\ \bibinfo {pages} {050201} (\bibinfo {year}
  {2014})}\BibitemShut {NoStop}%
\bibitem [{\citenamefont {Grob}\ \emph {et~al.}(2016)\citenamefont {Grob},
  \citenamefont {Zippelius},\ and\ \citenamefont {Heussinger}}]{grob_2016}%
  \BibitemOpen
  \bibfield  {author} {\bibinfo {author} {\bibfnamefont {M.}~\bibnamefont
  {Grob}}, \bibinfo {author} {\bibfnamefont {A.}~\bibnamefont {Zippelius}}, \
  and\ \bibinfo {author} {\bibfnamefont {C.}~\bibnamefont {Heussinger}},\
  }\bibfield  {title} {\enquote {\bibinfo {title} {Rheological chaos of
  frictional grains},}\ }\href@noop {} {\bibfield  {journal} {\bibinfo
  {journal} {Phys. Rev. E}\ }\textbf {\bibinfo {volume} {93}},\ \bibinfo
  {pages} {030901} (\bibinfo {year} {2016})}\BibitemShut {NoStop}%
\bibitem [{\citenamefont {Saw}\ \emph {et~al.}(2019)\citenamefont {Saw},
  \citenamefont {Grob}, \citenamefont {Zippelius},\ and\ \citenamefont
  {Heussinger}}]{saw_2019}%
  \BibitemOpen
  \bibfield  {author} {\bibinfo {author} {\bibfnamefont {S.}~\bibnamefont
  {Saw}}, \bibinfo {author} {\bibfnamefont {M.}~\bibnamefont {Grob}}, \bibinfo
  {author} {\bibfnamefont {A.}~\bibnamefont {Zippelius}}, \ and\ \bibinfo
  {author} {\bibfnamefont {C.}~\bibnamefont {Heussinger}},\ }\bibfield  {title}
  {\enquote {\bibinfo {title} {Unsteady flow, clusters and bands in a model
  shear-thickening fluid},}\ }\href@noop {} {\bibfield  {journal} {\bibinfo
  {journal} {arXiv preprint arXiv:1905.06174}\ } (\bibinfo {year}
  {2019})}\BibitemShut {NoStop}%
\bibitem [{\citenamefont {Holmes}\ \emph {et~al.}(2005)\citenamefont {Holmes},
  \citenamefont {Cates}, \citenamefont {Fuchs},\ and\ \citenamefont
  {Sollich}}]{holmes_2005}%
  \BibitemOpen
  \bibfield  {author} {\bibinfo {author} {\bibfnamefont {C.~B.}\ \bibnamefont
  {Holmes}}, \bibinfo {author} {\bibfnamefont {M.~E.}\ \bibnamefont {Cates}},
  \bibinfo {author} {\bibfnamefont {M.}~\bibnamefont {Fuchs}}, \ and\ \bibinfo
  {author} {\bibfnamefont {P.}~\bibnamefont {Sollich}},\ }\bibfield  {title}
  {\enquote {\bibinfo {title} {Glass transitions and shear thickening
  suspension rheology},}\ }\href@noop {} {\bibfield  {journal} {\bibinfo
  {journal} {J. Rheol.}\ }\textbf {\bibinfo {volume} {49}},\ \bibinfo {pages}
  {237--269} (\bibinfo {year} {2005})}\BibitemShut {NoStop}%
\bibitem [{\citenamefont {Thomas}\ \emph {et~al.}(2018)\citenamefont {Thomas},
  \citenamefont {Ramola}, \citenamefont {Singh}, \citenamefont {Mari},
  \citenamefont {Morris},\ and\ \citenamefont {Chakraborty}}]{thomas_2018}%
  \BibitemOpen
  \bibfield  {author} {\bibinfo {author} {\bibfnamefont {J.~E.}\ \bibnamefont
  {Thomas}}, \bibinfo {author} {\bibfnamefont {K.}~\bibnamefont {Ramola}},
  \bibinfo {author} {\bibfnamefont {A.}~\bibnamefont {Singh}}, \bibinfo
  {author} {\bibfnamefont {R.}~\bibnamefont {Mari}}, \bibinfo {author}
  {\bibfnamefont {J.~F.}\ \bibnamefont {Morris}}, \ and\ \bibinfo {author}
  {\bibfnamefont {B.}~\bibnamefont {Chakraborty}},\ }\bibfield  {title}
  {\enquote {\bibinfo {title} {Microscopic origin of frictional rheology in
  dense suspensions: correlations in force space},}\ }\href@noop {} {\bibfield
  {journal} {\bibinfo  {journal} {Phys. Rev. Lett.}\ }\textbf {\bibinfo
  {volume} {121}},\ \bibinfo {pages} {128002} (\bibinfo {year}
  {2018})}\BibitemShut {NoStop}%
\bibitem [{\citenamefont {Seifert}(2012)}]{seifert_2012}%
  \BibitemOpen
  \bibfield  {author} {\bibinfo {author} {\bibfnamefont {U.}~\bibnamefont
  {Seifert}},\ }\bibfield  {title} {\enquote {\bibinfo {title} {Stochastic
  thermodynamics, fluctuation theorems and molecular machines},}\ }\href@noop
  {} {\bibfield  {journal} {\bibinfo  {journal} {Rep. Prog. Phys.}\ }\textbf
  {\bibinfo {volume} {75}},\ \bibinfo {pages} {126001} (\bibinfo {year}
  {2012})}\BibitemShut {NoStop}%
\bibitem [{\citenamefont {Sekimoto}(2010)}]{sekimoto_2010}%
  \BibitemOpen
  \bibfield  {author} {\bibinfo {author} {\bibfnamefont {K.}~\bibnamefont
  {Sekimoto}},\ }\href@noop {} {\emph {\bibinfo {title} {Stochastic
  energetics}}},\ Vol.\ \bibinfo {volume} {799}\ (\bibinfo  {publisher}
  {Springer},\ \bibinfo {year} {2010})\BibitemShut {NoStop}%
\bibitem [{\citenamefont {Rahbari}\ \emph {et~al.}(2017)\citenamefont
  {Rahbari}, \citenamefont {Saberi}, \citenamefont {Park},\ and\ \citenamefont
  {Vollmer}}]{rahbari_2017}%
  \BibitemOpen
  \bibfield  {author} {\bibinfo {author} {\bibfnamefont {S.~H.~E.}\
  \bibnamefont {Rahbari}}, \bibinfo {author} {\bibfnamefont {A.~A.}\
  \bibnamefont {Saberi}}, \bibinfo {author} {\bibfnamefont {H.}~\bibnamefont
  {Park}}, \ and\ \bibinfo {author} {\bibfnamefont {J.}~\bibnamefont
  {Vollmer}},\ }\bibfield  {title} {\enquote {\bibinfo {title} {Characterizing
  rare fluctuations in soft particulate flows},}\ }\href@noop {} {\bibfield
  {journal} {\bibinfo  {journal} {Nat. Commun.}\ }\textbf {\bibinfo {volume}
  {8}},\ \bibinfo {pages} {11} (\bibinfo {year} {2017})}\BibitemShut {NoStop}%
\bibitem [{\citenamefont {Zheng}\ \emph {et~al.}(2018)\citenamefont {Zheng},
  \citenamefont {Sun}, \citenamefont {Wang},\ and\ \citenamefont
  {Zhang}}]{zheng_2018_2}%
  \BibitemOpen
  \bibfield  {author} {\bibinfo {author} {\bibfnamefont {J.}~\bibnamefont
  {Zheng}}, \bibinfo {author} {\bibfnamefont {A.}~\bibnamefont {Sun}}, \bibinfo
  {author} {\bibfnamefont {Y.}~\bibnamefont {Wang}}, \ and\ \bibinfo {author}
  {\bibfnamefont {J.}~\bibnamefont {Zhang}},\ }\bibfield  {title} {\enquote
  {\bibinfo {title} {Energy fluctuations in slowly sheared granular
  materials},}\ }\href@noop {} {\bibfield  {journal} {\bibinfo  {journal}
  {Phys. Rev. Lett.}\ }\textbf {\bibinfo {volume} {121}},\ \bibinfo {pages}
  {248001} (\bibinfo {year} {2018})}\BibitemShut {NoStop}%
\bibitem [{\citenamefont {Bos}\ and\ \citenamefont
  {Zamansky}(2019)}]{bos_2019}%
  \BibitemOpen
  \bibfield  {author} {\bibinfo {author} {\bibfnamefont {W.~J.~T.}\
  \bibnamefont {Bos}}\ and\ \bibinfo {author} {\bibfnamefont {Remi}\
  \bibnamefont {Zamansky}},\ }\bibfield  {title} {\enquote {\bibinfo {title}
  {Power fluctuations in turbulence},}\ }\href@noop {} {\bibfield  {journal}
  {\bibinfo  {journal} {Phys. Rev. Lett.}\ }\textbf {\bibinfo {volume} {122}},\
  \bibinfo {pages} {124504} (\bibinfo {year} {2019})}\BibitemShut {NoStop}%
\bibitem [{\citenamefont {Gerloff}\ and\ \citenamefont
  {Klapp}(2018)}]{gerloff_2018}%
  \BibitemOpen
  \bibfield  {author} {\bibinfo {author} {\bibfnamefont {S.}~\bibnamefont
  {Gerloff}}\ and\ \bibinfo {author} {\bibfnamefont {S.~H.~L.}\ \bibnamefont
  {Klapp}},\ }\bibfield  {title} {\enquote {\bibinfo {title} {Stochastic
  thermodynamics of a confined colloidal suspension under shear flow},}\
  }\href@noop {} {\bibfield  {journal} {\bibinfo  {journal} {Phys. Rev. E}\
  }\textbf {\bibinfo {volume} {98}},\ \bibinfo {pages} {062619} (\bibinfo
  {year} {2018})}\BibitemShut {NoStop}%
\bibitem [{\citenamefont {Makse}\ and\ \citenamefont
  {Kurchan}(2002)}]{makse_2002}%
  \BibitemOpen
  \bibfield  {author} {\bibinfo {author} {\bibfnamefont {H.~A.}\ \bibnamefont
  {Makse}}\ and\ \bibinfo {author} {\bibfnamefont {J.}~\bibnamefont
  {Kurchan}},\ }\bibfield  {title} {\enquote {\bibinfo {title} {Testing the
  thermodynamic approach to granular matter with a numerical model of a
  decisive experiment},}\ }\href@noop {} {\bibfield  {journal} {\bibinfo
  {journal} {Nature}\ }\textbf {\bibinfo {volume} {415}},\ \bibinfo {pages}
  {614--614} (\bibinfo {year} {2002})}\BibitemShut {NoStop}%
\bibitem [{\citenamefont {Goldhirsch}(2010)}]{goldhirsch_2010}%
  \BibitemOpen
  \bibfield  {author} {\bibinfo {author} {\bibfnamefont {I.}~\bibnamefont
  {Goldhirsch}},\ }\bibfield  {title} {\enquote {\bibinfo {title} {Stress,
  stress asymmetry and couple stress: from discrete particles to continuous
  fields},}\ }\href@noop {} {\bibfield  {journal} {\bibinfo  {journal}
  {Granular Matter}\ }\textbf {\bibinfo {volume} {12}},\ \bibinfo {pages}
  {239--252} (\bibinfo {year} {2010})}\BibitemShut {NoStop}%
\bibitem [{\citenamefont {Mitarai}\ \emph {et~al.}(2002)\citenamefont
  {Mitarai}, \citenamefont {Hayakawa},\ and\ \citenamefont
  {Nakanishi}}]{mitarai_2002}%
  \BibitemOpen
  \bibfield  {author} {\bibinfo {author} {\bibfnamefont {N.}~\bibnamefont
  {Mitarai}}, \bibinfo {author} {\bibfnamefont {H.}~\bibnamefont {Hayakawa}}, \
  and\ \bibinfo {author} {\bibfnamefont {H.}~\bibnamefont {Nakanishi}},\
  }\bibfield  {title} {\enquote {\bibinfo {title} {Collisional granular flow as
  a micropolar fluid},}\ }\href@noop {} {\bibfield  {journal} {\bibinfo
  {journal} {Phys. Rev. Lett.}\ }\textbf {\bibinfo {volume} {88}},\ \bibinfo
  {pages} {174301} (\bibinfo {year} {2002})}\BibitemShut {NoStop}%
\bibitem [{\citenamefont {Cheng}\ \emph {et~al.}(2011)\citenamefont {Cheng},
  \citenamefont {McCoy}, \citenamefont {Israelachvili},\ and\ \citenamefont
  {Cohen}}]{cheng_2011}%
  \BibitemOpen
  \bibfield  {author} {\bibinfo {author} {\bibfnamefont {X.}~\bibnamefont
  {Cheng}}, \bibinfo {author} {\bibfnamefont {J.~H}\ \bibnamefont {McCoy}},
  \bibinfo {author} {\bibfnamefont {J.~N.}\ \bibnamefont {Israelachvili}}, \
  and\ \bibinfo {author} {\bibfnamefont {I.}~\bibnamefont {Cohen}},\ }\bibfield
   {title} {\enquote {\bibinfo {title} {Imaging the microscopic structure of
  shear thinning and thickening colloidal suspensions},}\ }\href@noop {}
  {\bibfield  {journal} {\bibinfo  {journal} {Science}\ }\textbf {\bibinfo
  {volume} {333}},\ \bibinfo {pages} {1276--1279} (\bibinfo {year}
  {2011})}\BibitemShut {NoStop}%
\bibitem [{\citenamefont {Hoffman}(1974)}]{hoffman_1974}%
  \BibitemOpen
  \bibfield  {author} {\bibinfo {author} {\bibfnamefont {R.~L.}\ \bibnamefont
  {Hoffman}},\ }\bibfield  {title} {\enquote {\bibinfo {title} {Discontinuous
  and dilatant viscosity behavior in concentrated suspensions. ii. theory and
  experimental tests},}\ }\href@noop {} {\bibfield  {journal} {\bibinfo
  {journal} {J. Colloid Interface Sci.}\ }\textbf {\bibinfo {volume} {46}},\
  \bibinfo {pages} {491--506} (\bibinfo {year} {1974})}\BibitemShut {NoStop}%
\bibitem [{\citenamefont {Hoffman}(1998)}]{hoffman_1998}%
  \BibitemOpen
  \bibfield  {author} {\bibinfo {author} {\bibfnamefont {R.~L.}\ \bibnamefont
  {Hoffman}},\ }\bibfield  {title} {\enquote {\bibinfo {title} {Explanations
  for the cause of shear thickening in concentrated colloidal suspensions},}\
  }\href@noop {} {\bibfield  {journal} {\bibinfo  {journal} {J. Rheol.}\
  }\textbf {\bibinfo {volume} {42}},\ \bibinfo {pages} {111--123} (\bibinfo
  {year} {1998})}\BibitemShut {NoStop}%
\bibitem [{\citenamefont {Chattoraj}\ \emph
  {et~al.}(2019{\natexlab{a}})\citenamefont {Chattoraj}, \citenamefont
  {Gendelman}, \citenamefont {Ciamarra},\ and\ \citenamefont
  {Procaccia}}]{chattoraj_2019}%
  \BibitemOpen
  \bibfield  {author} {\bibinfo {author} {\bibfnamefont {J.}~\bibnamefont
  {Chattoraj}}, \bibinfo {author} {\bibfnamefont {O.}~\bibnamefont
  {Gendelman}}, \bibinfo {author} {\bibfnamefont {M.~P.}\ \bibnamefont
  {Ciamarra}}, \ and\ \bibinfo {author} {\bibfnamefont {I.}~\bibnamefont
  {Procaccia}},\ }\bibfield  {title} {\enquote {\bibinfo {title} {Noise
  amplification in frictional systems: Oscillatory instabilities},}\
  }\href@noop {} {\bibfield  {journal} {\bibinfo  {journal} {arXiv preprint
  arXiv:1903.10887}\ } (\bibinfo {year} {2019}{\natexlab{a}})}\BibitemShut
  {NoStop}%
\bibitem [{\citenamefont {Chattoraj}\ \emph
  {et~al.}(2019{\natexlab{b}})\citenamefont {Chattoraj}, \citenamefont
  {Gendelman}, \citenamefont {Ciamarra},\ and\ \citenamefont
  {Procaccia}}]{chattoraj_2019_2}%
  \BibitemOpen
  \bibfield  {author} {\bibinfo {author} {\bibfnamefont {J.}~\bibnamefont
  {Chattoraj}}, \bibinfo {author} {\bibfnamefont {O.}~\bibnamefont
  {Gendelman}}, \bibinfo {author} {\bibfnamefont {M.~P.}\ \bibnamefont
  {Ciamarra}}, \ and\ \bibinfo {author} {\bibfnamefont {I.}~\bibnamefont
  {Procaccia}},\ }\bibfield  {title} {\enquote {\bibinfo {title} {Oscillatory
  instabilities in frictional granular matter},}\ }\href@noop {} {\bibfield
  {journal} {\bibinfo  {journal} {Phys. Rev. Lett.}\ }\textbf {\bibinfo
  {volume} {123}} (\bibinfo {year} {2019}{\natexlab{b}})}\BibitemShut {NoStop}%
\bibitem [{\citenamefont {Maloney}\ and\ \citenamefont
  {Lemaitre}(2004)}]{maloney_2004}%
  \BibitemOpen
  \bibfield  {author} {\bibinfo {author} {\bibfnamefont {C.}~\bibnamefont
  {Maloney}}\ and\ \bibinfo {author} {\bibfnamefont {A.}~\bibnamefont
  {Lemaitre}},\ }\bibfield  {title} {\enquote {\bibinfo {title} {Universal
  breakdown of elasticity at the onset of material failure},}\ }\href@noop {}
  {\bibfield  {journal} {\bibinfo  {journal} {Phys. Rev. Lett.}\ }\textbf
  {\bibinfo {volume} {93}},\ \bibinfo {pages} {195501} (\bibinfo {year}
  {2004})}\BibitemShut {NoStop}%
\bibitem [{\citenamefont {Felzer}\ and\ \citenamefont
  {Brodsky}(2006)}]{felzer_2006}%
  \BibitemOpen
  \bibfield  {author} {\bibinfo {author} {\bibfnamefont {K.~R.}\ \bibnamefont
  {Felzer}}\ and\ \bibinfo {author} {\bibfnamefont {E.~E.}\ \bibnamefont
  {Brodsky}},\ }\bibfield  {title} {\enquote {\bibinfo {title} {Decay of
  aftershock density with distance indicates triggering by dynamic stress},}\
  }\href@noop {} {\bibfield  {journal} {\bibinfo  {journal} {Nature}\ }\textbf
  {\bibinfo {volume} {441}},\ \bibinfo {pages} {735} (\bibinfo {year}
  {2006})}\BibitemShut {NoStop}%
\bibitem [{\citenamefont {Kardar}(2007)}]{kardar_2007}%
  \BibitemOpen
  \bibfield  {author} {\bibinfo {author} {\bibfnamefont {M.}~\bibnamefont
  {Kardar}},\ }\href@noop {} {\emph {\bibinfo {title} {Statistical physics of
  fields}}}\ (\bibinfo  {publisher} {Cambridge University Press},\ \bibinfo
  {year} {2007})\BibitemShut {NoStop}%
\bibitem [{\citenamefont {Edwards}\ and\ \citenamefont
  {Oakeshott}(1989)}]{edwards_1989}%
  \BibitemOpen
  \bibfield  {author} {\bibinfo {author} {\bibfnamefont {S.~F.}\ \bibnamefont
  {Edwards}}\ and\ \bibinfo {author} {\bibfnamefont {R.~B.~S.}\ \bibnamefont
  {Oakeshott}},\ }\bibfield  {title} {\enquote {\bibinfo {title} {Theory of
  powders},}\ }\href@noop {} {\bibfield  {journal} {\bibinfo  {journal}
  {Physica A}\ }\textbf {\bibinfo {volume} {157}},\ \bibinfo {pages} {1080}
  (\bibinfo {year} {1989})}\BibitemShut {NoStop}%
\bibitem [{\citenamefont {Bi}\ \emph {et~al.}(2015)\citenamefont {Bi},
  \citenamefont {Henkes}, \citenamefont {Daniels},\ and\ \citenamefont
  {Chakraborty}}]{bi_2015_2}%
  \BibitemOpen
  \bibfield  {author} {\bibinfo {author} {\bibfnamefont {D.}~\bibnamefont
  {Bi}}, \bibinfo {author} {\bibfnamefont {S.}~\bibnamefont {Henkes}}, \bibinfo
  {author} {\bibfnamefont {K.~E.}\ \bibnamefont {Daniels}}, \ and\ \bibinfo
  {author} {\bibfnamefont {B.}~\bibnamefont {Chakraborty}},\ }\bibfield
  {title} {\enquote {\bibinfo {title} {The statistical physics of athermal
  materials},}\ }\href@noop {} {\bibfield  {journal} {\bibinfo  {journal}
  {Annu. Rev. Condens. Matter Phys.}\ }\textbf {\bibinfo {volume} {6}},\
  \bibinfo {pages} {63--83} (\bibinfo {year} {2015})}\BibitemShut {NoStop}%
\bibitem [{\citenamefont {Dean}\ and\ \citenamefont
  {Lefevre}(2003)}]{dean_2003}%
  \BibitemOpen
  \bibfield  {author} {\bibinfo {author} {\bibfnamefont {D.~S}\ \bibnamefont
  {Dean}}\ and\ \bibinfo {author} {\bibfnamefont {A.}~\bibnamefont {Lefevre}},\
  }\bibfield  {title} {\enquote {\bibinfo {title} {Possible test of the
  thermodynamic approach to granular media},}\ }\href@noop {} {\bibfield
  {journal} {\bibinfo  {journal} {Phys. Rev. Lett.}\ }\textbf {\bibinfo
  {volume} {90}},\ \bibinfo {pages} {198301} (\bibinfo {year}
  {2003})}\BibitemShut {NoStop}%
\bibitem [{\citenamefont {Zhao}\ \emph {et~al.}(2012)\citenamefont {Zhao},
  \citenamefont {Sidle}, \citenamefont {Swinney},\ and\ \citenamefont
  {Schr{\"o}ter}}]{zhao_2012}%
  \BibitemOpen
  \bibfield  {author} {\bibinfo {author} {\bibfnamefont {S.~C.}\ \bibnamefont
  {Zhao}}, \bibinfo {author} {\bibfnamefont {S.}~\bibnamefont {Sidle}},
  \bibinfo {author} {\bibfnamefont {H.~L.}\ \bibnamefont {Swinney}}, \ and\
  \bibinfo {author} {\bibfnamefont {M.}~\bibnamefont {Schr{\"o}ter}},\
  }\bibfield  {title} {\enquote {\bibinfo {title} {Correlation between voronoi
  volumes in disc packings},}\ }\href@noop {} {\bibfield  {journal} {\bibinfo
  {journal} {Europhys. Lett.}\ }\textbf {\bibinfo {volume} {97}},\ \bibinfo
  {pages} {34004} (\bibinfo {year} {2012})}\BibitemShut {NoStop}%
\bibitem [{\citenamefont {McNamara}\ \emph {et~al.}(2009)\citenamefont
  {McNamara}, \citenamefont {Richard}, \citenamefont {De~Richter},
  \citenamefont {Le~Ca{\"e}r},\ and\ \citenamefont {Delannay}}]{mcnamara_2009}%
  \BibitemOpen
  \bibfield  {author} {\bibinfo {author} {\bibfnamefont {S.}~\bibnamefont
  {McNamara}}, \bibinfo {author} {\bibfnamefont {P.}~\bibnamefont {Richard}},
  \bibinfo {author} {\bibfnamefont {S.~K.}\ \bibnamefont {De~Richter}},
  \bibinfo {author} {\bibfnamefont {G.}~\bibnamefont {Le~Ca{\"e}r}}, \ and\
  \bibinfo {author} {\bibfnamefont {R.}~\bibnamefont {Delannay}},\ }\bibfield
  {title} {\enquote {\bibinfo {title} {Measurement of granular entropy},}\
  }\href@noop {} {\bibfield  {journal} {\bibinfo  {journal} {Phys. Rev. E}\
  }\textbf {\bibinfo {volume} {80}},\ \bibinfo {pages} {031301} (\bibinfo
  {year} {2009})}\BibitemShut {NoStop}%
\bibitem [{\citenamefont {Bililign}\ \emph {et~al.}(2019)\citenamefont
  {Bililign}, \citenamefont {Kollmer},\ and\ \citenamefont
  {Daniels}}]{bililign_2019}%
  \BibitemOpen
  \bibfield  {author} {\bibinfo {author} {\bibfnamefont {E.~S.}\ \bibnamefont
  {Bililign}}, \bibinfo {author} {\bibfnamefont {J.~E.}\ \bibnamefont
  {Kollmer}}, \ and\ \bibinfo {author} {\bibfnamefont {K.~E.}\ \bibnamefont
  {Daniels}},\ }\bibfield  {title} {\enquote {\bibinfo {title} {Protocol
  dependence and state variables in the force-moment ensemble},}\ }\href@noop
  {} {\bibfield  {journal} {\bibinfo  {journal} {Phys. Rev. Lett.}\ }\textbf
  {\bibinfo {volume} {122}},\ \bibinfo {pages} {038001} (\bibinfo {year}
  {2019})}\BibitemShut {NoStop}%
\bibitem [{\citenamefont {Behringer}\ \emph {et~al.}(2008)\citenamefont
  {Behringer}, \citenamefont {Bi}, \citenamefont {Chakraborty}, \citenamefont
  {Henkes},\ and\ \citenamefont {Hartley}}]{behringer_2008}%
  \BibitemOpen
  \bibfield  {author} {\bibinfo {author} {\bibfnamefont {R.~P.}\ \bibnamefont
  {Behringer}}, \bibinfo {author} {\bibfnamefont {D.}~\bibnamefont {Bi}},
  \bibinfo {author} {\bibfnamefont {B.}~\bibnamefont {Chakraborty}}, \bibinfo
  {author} {\bibfnamefont {S.}~\bibnamefont {Henkes}}, \ and\ \bibinfo {author}
  {\bibfnamefont {R.~R.}\ \bibnamefont {Hartley}},\ }\bibfield  {title}
  {\enquote {\bibinfo {title} {Why do granular materials stiffen with shear
  rate? test of novel stress-based statistics},}\ }\href@noop {} {\bibfield
  {journal} {\bibinfo  {journal} {Phys. Rev. Lett.}\ }\textbf {\bibinfo
  {volume} {101}},\ \bibinfo {pages} {268301} (\bibinfo {year}
  {2008})}\BibitemShut {NoStop}%
\bibitem [{\citenamefont {Poeschel}\ and\ \citenamefont
  {Schwager}(2005)}]{poeschel2005}%
  \BibitemOpen
  \bibfield  {author} {\bibinfo {author} {\bibfnamefont {T.}~\bibnamefont
  {Poeschel}}\ and\ \bibinfo {author} {\bibfnamefont {T.}~\bibnamefont
  {Schwager}},\ }\href@noop {} {\emph {\bibinfo {title} {Computational granular
  dynamics}}}\ (\bibinfo  {publisher} {Springer},\ \bibinfo {year}
  {2005})\BibitemShut {NoStop}%
\bibitem [{\citenamefont {van Hecke}(2010)}]{vanhecke_2010}%
  \BibitemOpen
  \bibfield  {author} {\bibinfo {author} {\bibfnamefont {M}~\bibnamefont {van
  Hecke}},\ }\bibfield  {title} {\enquote {\bibinfo {title} {Jamming of soft
  particles: geometry, mechanics, scaling and isostaticity},}\ }\href@noop {}
  {\bibfield  {journal} {\bibinfo  {journal} {J. Phys.: Condens. Matter}\
  }\textbf {\bibinfo {volume} {22}},\ \bibinfo {pages} {033101} (\bibinfo
  {year} {2010})}\BibitemShut {NoStop}%
\bibitem [{\citenamefont {Bi}\ \emph {et~al.}(2011)\citenamefont {Bi},
  \citenamefont {Zhang}, \citenamefont {Chakraborty},\ and\ \citenamefont
  {Behringer}}]{bi_2011}%
  \BibitemOpen
  \bibfield  {author} {\bibinfo {author} {\bibfnamefont {D.}~\bibnamefont
  {Bi}}, \bibinfo {author} {\bibfnamefont {J.}~\bibnamefont {Zhang}}, \bibinfo
  {author} {\bibfnamefont {B.}~\bibnamefont {Chakraborty}}, \ and\ \bibinfo
  {author} {\bibfnamefont {R.~P.}\ \bibnamefont {Behringer}},\ }\bibfield
  {title} {\enquote {\bibinfo {title} {Jamming by shear},}\ }\href@noop {}
  {\bibfield  {journal} {\bibinfo  {journal} {Nature}\ }\textbf {\bibinfo
  {volume} {480}},\ \bibinfo {pages} {355} (\bibinfo {year}
  {2011})}\BibitemShut {NoStop}%
\end{thebibliography}
